\documentclass[aps,prd,nofootinbib,twocolumn,superscriptaddress,preprintnumbers]{revtex4-1}

\usepackage{amssymb}
\usepackage{amsmath}
\usepackage{epsfig}
\usepackage{hyperref}
\usepackage{breakurl}
\usepackage{bm}
\usepackage{color}
\usepackage{tikz}
\usepackage[utf8]{inputenc}
\usetikzlibrary{decorations.markings}

\makeatletter
\def\simgt{\mathrel{\lower2.5pt\vbox{\lineskip=0pt\baselineskip=0pt
           \hbox{$>$}\hbox{$\sim$}}}}
\def\simlt{\mathrel{\lower2.5pt\vbox{\lineskip=0pt\baselineskip=0pt
           \hbox{$<$}\hbox{$\sim$}}}}

\makeatother

\def\spa#1.#2{\left\langle#1\,#2\right\rangle}
\def\spb#1.#2{\left[#1\,#2\right]}
\def\sand#1.#2.#3{%
\left\langle#1{\vphantom1}\right|{#2}\left|#3\right]}%
\def\sandmp#1.#2.#3{%
\left\langle#1{\vphantom1}\right|{#2}\left|#3\right]}%
\def\sandpm#1.#2.#3{%
\left[#1{\vphantom1}\right|{#2}\left|#3\right\rangle}%
\def\sandmm#1.#2.#3{%
\left\langle#1{\vphantom1}\right|{#2}\left|#3\right\rangle}%
\def\sandpp#1.#2.#3{%
\left[#1{\vphantom1}\right|{#2}\left|#3\right]}%

\def\nn{\nonumber}
\def\Section#1{\noindent {\it #1}}
\newcommand{\be}{\begin{equation}}
\newcommand{\ee}{\end{equation}}
\newcommand{\eq}[2]{\be\begin{aligned}#1 \label{#2}\end{aligned}\ee}

\newcommand{\Eq}[1]{Eq.~\eqref{#1}}
\newcommand{\Eqs}[2]{Eqs.~\eqref{#1} and \eqref{#2}}

\newcommand{\E}{{\rm E}}
\newcommand{\K}{{\rm K}}

\def\topbotatom#1{\hbox{\hbox to 0pt{$#1\bot$\hss}$#1\top$}}

\newcommand{\tabeq}[2]{ \parbox{#1}{  \be\begin{aligned}#2 \end{aligned} \nonumber \ee }}

\begin{document}
\preprint{\preprint{CALT-TH-2021-004, FR-PHENO-2021-03, OUTP-21-03P}}
\title{Scattering Amplitudes and Conservative Binary Dynamics at ${\cal O}(G^4)$}

\author{Zvi Bern}
\affiliation{
Mani L. Bhaumik Institute for Theoretical Physics,
University of California at Los Angeles,
Los Angeles, CA 90095, USA}
\author{Julio Parra-Martinez}
\affiliation{Walter Burke Institute for Theoretical Physics,
    California Institute of Technology, Pasadena, CA 91125}
\author{Radu Roiban}
\affiliation{Institute for Gravitation and the Cosmos,
Pennsylvania State University,
University Park, Pz 16802, USA}
\author{Michael~S.~Ruf}
\affiliation{Physikalisches Institut, Albert-Ludwigs-Universit\"at Freiburg, 
Hermann-Herder-Strasse 3, 79104 Freiburg, Germany}
\author{Chia-Hsien Shen}
\affiliation{
Department of Physics, University of California at San Diego, 9500 Gilman Drive, La Jolla, CA 92093-0319, USA}
\author{ Mikhail P. Solon}
\affiliation{
Mani L. Bhaumik Institute for Theoretical Physics,
University of California at Los Angeles,
Los Angeles, CA 90095, USA}
\author{Mao Zeng}
\affiliation{Rudolf Peierls Centre for Theoretical Physics, University of Oxford, Parks Road, Oxford OX1 3PU, United Kingdom}

\begin{abstract}
Using scattering amplitudes, we obtain the potential contributions to
conservative binary dynamics in general relativity at fourth
post-Minkowskian order, ${\cal O}(G^4)$.  As in previous lower-order
calculations, we harness powerful tools from the modern scattering
amplitudes program including generalized unitarity, the double copy,
and advanced multiloop integration methods, in combination with
effective field theory. The classical amplitude involves polylogarithms with up
to transcendental weight two and elliptic integrals.  We derive the
radial action directly from the amplitude, and determine the corresponding Hamiltonian in isotropic
gauge. Our results are in agreement with
known overlapping terms up to sixth post-Newtonian order, and
with the probe limit. We also determine the post-Minkowskian energy loss from radiation emission at ${\cal O}(G^3)$ via its relation to the tail effect.
\end{abstract}

\maketitle

\Section{Introduction.}
The emergence of gravitational-wave science~\cite{LIGO} has
dramatically underscored the scientific value of observing the
universe through an entirely new lens, and will continue to
fundamentally transform key areas in astronomy, cosmology, and
particle physics. This calls for invigorating the theoretical
framework necessary for interpreting signals at current and future
detectors~\cite{NewDetectors}, and has thus galvanized new work in
this direction.  This includes a new program~\cite{2PM,3PM,3PMLong}
for understanding the nature of gravitational-wave sources based on
tools from scattering amplitudes and effective field theory
(EFT). The connection of scattering amplitudes to general relativity
corrections to Newton's potential has long been
known~\cite{ScatteringToGrav,RecentAmpToPotential,NeillandRothstein}.
Starting from foundational ideas from EFT applied to
gravitational-wave physics~\cite{NRGR}, this new effort has integrated generalized unitarity~\cite{GeneralizedUnitarity},
double-copy relations between gauge and gravity
theories~\cite{KLT,BCJ}, EFT~\cite{2PM,NeillandRothstein}, and advanced multiloop
integration~\cite{IBP,DEs,FIRE}. These ideas culminated with
advancing the state of the art by obtaining the ${\cal O}(G^3)$
conservative Hamiltonian for spinless compact
binaries~\cite{3PM,3PMLong}, whose various aspects have now been
confirmed in multiple studies~\cite{6PNBlumlein,3PMFeynman,3PMworldline,Bini:2020wpo,Bini:2020nsb}.  In this paper, we
extend previous methods and take the next step to obtain the
contributions to conservative binary dynamics at ${\cal O}(G^4)$.

We focus here on the inspiral phase of conservative binary dynamics.
Traditionally, this is approached using effective one-body~\cite{EOB},
numerical relativity~\cite{NR}, gravitational
self-force~\cite{self_force,SelfForceReview}, and perturbation theory in the
post-Newtonian (PN)~\cite{PN,4PNDJS, 4PNrest, Bini:2017wfr}, post-Minkowskian
(PM)~\cite{PM,DamourPM}, and non-relativistic general relativity
(NRGR)~\cite{NRGR,NRGRrecent} frameworks.  The PM approach has recently risen
in prominence due to increased analytic
control~\cite{2PM,3PM,3PMLong,DamourPM,NBI,PMModernList, Kosower:2018adc,
Paolo2PM, B2B}. We work in this context given the natural fit with relativistic
amplitudes. 

The new amplitudes-based approach has, of course, benefited immensely
from traditional methods, both in guidance and for confirming
calculations.  In turn, it has revealed connections between scattering
amplitudes, classical observables, and gravitational self-force,
inspiring new methods for obtaining perturbative
corrections~\cite{DamourSelfForce, B2B,SpinTutiFruiti,Bini:2020wpo,Bini:2020nsb,Bini:2020hmy}.

In this work we develop a new way to combine the techniques of
Refs.~\cite{2PM,3PM,3PMLong}, with advanced loop integration methods for
classical integrals introduced in Ref.~\cite{PRZ}, and a link between the
gauge-invariant amplitude and classical radial action, which further streamline
the amplitudes approach and demonstrate its scalability.  We derive the
scattering amplitude in the classical limit for two massive scalars interacting
via potential gravitons at ${\cal O}(G^4)$ and all orders in velocity. We also
obtain the corresponding conservative two-body Hamiltonian and radial action,
and determine the energy loss due to graviton emission at ${\cal O}(G^3)$
through its relation~\cite{Bini:2017wfr, BFLS, Bini:2020hmy} to the ${\cal
O}(G^4)$ tail effect~\cite{TailEffect,Galley}.  This order in perturbation
theory presents new features from the tail effect, which manifests an infrared
(IR) divergence due to the overlap between the momentum
regions~\cite{Beneke:1997zp} of potential and radiation
gravitons~\cite{Zerobin}. We also encounter a class of elliptic integrals,
which complicates the analysis.

Scattering amplitudes are independent of gauge or coordinate choices,
while EFT exposes universality in physical systems. These features
greatly help identify emergent structures that can enhance
our understanding of basic phenomena and lead to new tools that will
further the cycle of innovation.
Here we present a remarkably simple gauge-invariant relation between the conservative scattering
amplitude and the radial action based on a reorganization of the
amplitude into classical and iteration pieces, distinct from that of
Refs.~\cite{2PM,3PM,3PMLong,B2B}. It is well known that the radial action
is also gauge-invariant and encodes the dynamics of both bound and
unbound orbits (see e.g. Refs.~\cite{DamourRadial, Bini:2020hmy, B2B}). 

\Section{Classical Dynamics from Scattering Amplitudes.} 
We focus on conservative two-body dynamics for spinless compact objects,
described by the four-point amplitude,  ${\cal M}(\bm{q})$,  of gravitationally
interacting minimally-coupled massive scalars.
The two incoming particles of momenta $p_1, p_2$ have masses $m_1,m_2$,
and we define $\sigma \equiv { p_1 \cdot p_2 \over m_1 m_2}$ in mostly minus signature.
We work in the center-of-mass (COM) frame where the momentum transfer $q^\mu=(0,\bm q)$
is purely spatial.
Following Refs.~\cite{Landshoff:1989ig, PRZ}, we decompose $p_1, p_2$ into components orthogonal and along $q$, i.e., $p_{1} = \bar{p}_{1} - q/2$, $p_{2} =
\bar{p}_{2} + q/2$ with $\bar{p}_i \cdot q=0$.

As described in Refs.~\cite{2PM,3PM,3PMLong}, major simplifications are
obtained by taking the classical limit early at the level of the integrand. This
is achieved by an expansion in large angular momentum $J \gg \hbar$.  We implement this by rescaling $q, \ell \to \lambda q, \lambda
\ell$, where $\ell$ is any graviton momentum, and then
expanding in small $\lambda$.

The classical limit therefore identifies the soft region, defined by
the loop momentum scaling $\ell^\mu = (\omega, \bm \ell) \sim
(\lambda, \lambda)$, as encoding classical dynamics. 
In the spirit of
EFT~\cite{NRGR}, we simplify the analysis, especially in the presence
of the tail effect at ${\cal O}(G^4)$, by focusing on the potential and (ultrasoft) radiation subregions defined by
the scalings $\sim(v \lambda , \lambda)$ and $\sim(v \lambda, v
\lambda)$, respectively. Here and below we use $v$ to denote the typical velocity of the binary constituents, corresponding to the small velocity that defines
the PN expansion.

In the present work, we focus on the conservative part described by 
the potential contribution, and do
not include radiation. This is sufficient for completely specifying
the conservative dynamics through ${\cal
  O}(G^3)$~\cite{3PM,3PMLong}. However, at ${\cal O}(G^4)$, radiative
effects contribute to conservative dynamics via the tail
effect~\cite{TailEffect}. Since the potential and
radiation contributions overlap, this introduces
scheme-dependence and IR divergence~\cite{Zerobin}. We use conventional
dimensional regularization, where the amplitudes, including graviton
polarizations, are uniformly continued into $D=4-2\epsilon$
dimensions. 

%%%%%%%%%%%%%%%%%%%%%%%%%%%%%%%%%%%%
\Section{Amplitude-Action Relation.} 
Conservative binary dynamics 
is fully encoded in the four-point
amplitude ${\cal M}(\bm{q})$, truncated to the classical order.  
There
exists another scalar gauge invariant function which encodes the same
dynamics, namely the radial action,
which is defined as the integral of the radial momentum $p_r$ along the scattering trajectory,
$I_r(J) \equiv \int p_r dr$, with appropriate regularization of the long-distance contribution. Here we present a simple relation between these two
gauge-invariant quantities, exposed through the EFT introduced in Ref.~\cite{2PM}.

%%%%%%%%%%%%%% FIGURE %%%%%%%%%%%%%                             
\begin{figure}[tb]
	\begin{center}
		\includegraphics[scale=.26]{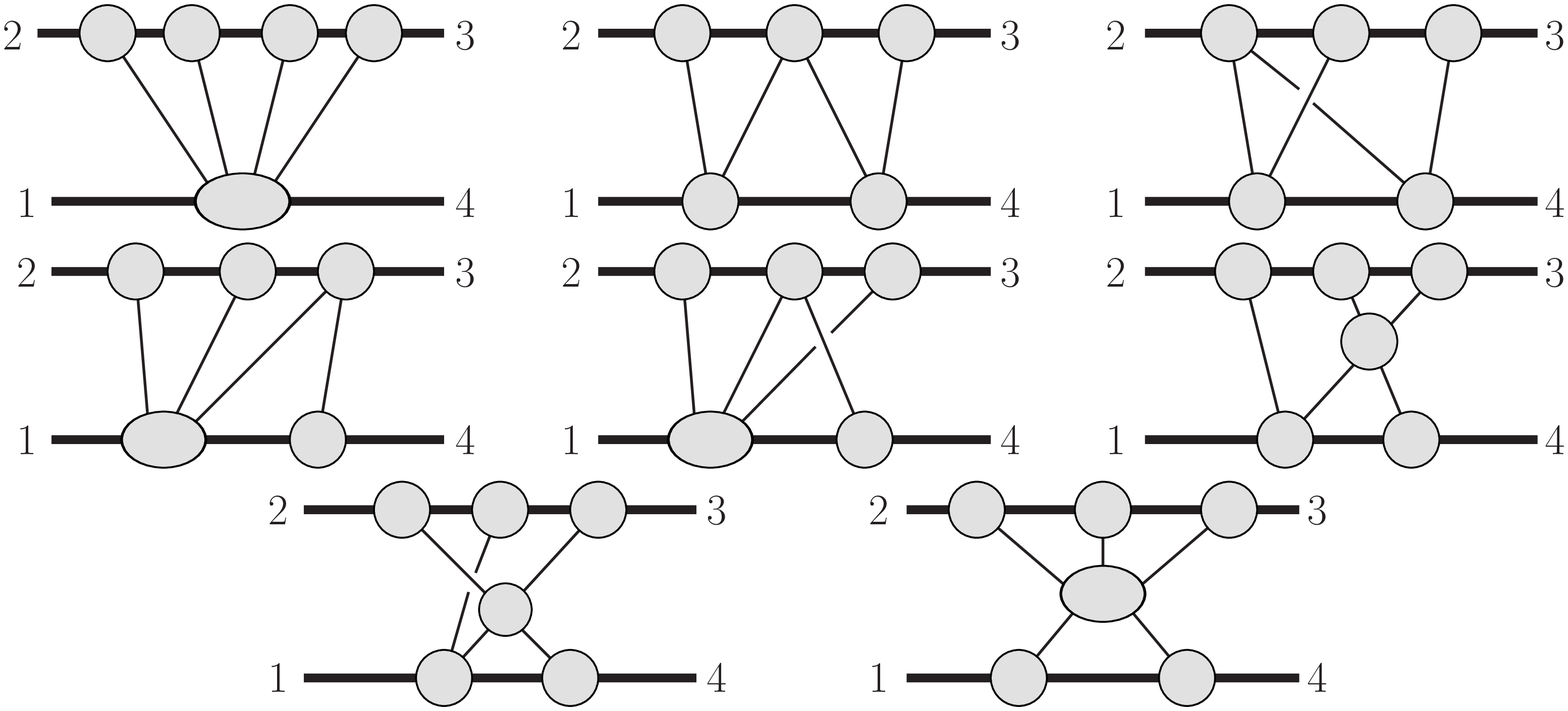}
	\end{center}
	\vskip -.5cm
	\caption{Generalized unitarity cuts encoding potential-region contributions to binary dynamics.
		Ovals represent tree amplitudes while exposed lines
		depict on-shell states.  Thin and thick lines denote
		gravitons and massive scalars, respectively.
	}
        \vspace{-0.5cm}
	\label{fig:cuts}
\end{figure}
%%%%%%%%%%%%%%%%%%%%%%%%%%%%%%%%%%%

In the classical limit, the amplitude at ${\cal O}(G^n)$ contains a classical
contribution that scales as $\lambda^{n-3}$ and iteration contributions that
scale as $\lambda^{n-2}, \lambda^{n-1}, \cdots, \lambda^{-2}$. The latter
correspond to iterations of lower-order amplitudes, are IR
divergent, and cancel in physical observables. 
Although the full amplitude is invariant, the choice of pole structure of the
iterations is not unique and the classical part is modified accordingly.
Previously this was chosen to
align with the matter energy poles in the EFT for direct cancellation without
explicit evaluation~\cite{2PM}. This choice also revealed a connection between the classical amplitude and the local COM momentum in isotropic gauge, first observed in~\cite{3PM,3PMLong} and later proven in~\cite{NBI,B2B}.
%This choice also had nice properties relating scattering amplitudes and classical quantities~\cite{3PM, 3PMLong, NBI, B2B}.

In the present analysis, we instead expand the matter poles about the momentum component along $\hat{\bm z}$, the direction of the spatial component of $\bar{p}_{1}$. 
Inspired by the eikonal approximation~\cite{EikonalPapers},
this prescription 
reveals a gauge-invariant ``amplitude-action relation''
\eq{
	i {\cal M}(\bm q) = \int_{J}\,\left( e^{i I_r(J)} - 1 \right)\,,
}{eq:action} 
between the amplitude ${\cal M}(\bm q)$ and the radial action $I_r(J)$.
The classical part of the amplitude then corresponds to the term linear in $I_r(J)$, given by
\begin{equation}
\tilde I_r (\bm q) = \int_{J} I_r(J) \equiv 4E|\bm p|\int \mu^{-2\epsilon} {d^{D-2}\bm b}\, e^{i\bm q\cdot \bm b}\, I_r(J) \,,
\label{eq:FT_J}
\end{equation}
where $\bm p$ is 
the spatial momentum, $E$ is the total energy, $|\bm b|=J/| \bm p|$ is the impact parameter in the COM frame,
and $\mu$ is the renormalization scale.
As will be shown elsewhere~\cite{4PMlong},
terms higher order in $I_r(J)$ in the relation \eqref{eq:action} have the following structure under our prescription
\begin{align}
&\int_{J} \frac{(iI_r(J))^n }{n!}
= i\int_{\bm \ell}  \frac{\tilde I_r(\bm{\ell}_1) \dots \tilde I_r(\bm{\ell}_n)}{Z_1 \dots Z_{n-1}} 
\,,
\label{eq:product} \\
&Z_j = - 4 E |\bm{p}| \big((\bm \ell_{1}+\bm \ell_2+\dots +\bm\ell_{j})\cdot \hat{\bm z}+i0 \big), \nn
\end{align}
where $\int_{\bm \ell} \equiv \int \prod_{i=1}^n \,\frac{d^{D-1}\bm \ell_i}{(2\pi)^{D-1}}\, (2\pi)^{D-1} \delta(\sum_{j=1}^n\bm\ell_j-\bm q)$ and we only keep the leading classical expansion in the numerator of \Eq{eq:product}.
Crucially, we manifest the pole structure in $Z_j$ when computing the amplitude such that the classical part can be isolated and iterations can be safely dropped without explicitly evaluating them, following the path of Ref.~\cite{2PM}.
With our prescription, we avoid tracking such terms, which is necessary in standard eikonal exponentiation~\cite{EikonalPapers}.

This amplitude-action relation then dictates the iteration structure in the amplitude when expanded in $G$. 
To illustrate, \Eqs{eq:action}{eq:product} with the expansions ${\cal M}(\bm q) = \sum_n {\cal M}_{n}(\bm q)$ and $\tilde{I}_r(\bm q) = \sum_n \tilde{I}_{r,n}(\bm q)$ yield
\begin{align}
	{\cal M}_1(\bm q) &= \tilde{I}_{r,1}(\bm q)\,, \qquad  
	{\cal M}_2(\bm q) = \tilde{I}_{r,2}(\bm q) + \int_{\bm \ell}\frac{\tilde{I}_{r,1} \tilde{I}_{r,1}}{Z_1} \,, \nonumber  \\
	{\cal M}_3(\bm q) &= \tilde{I}_{r,3}(\bm q) + \int_{\bm \ell}\frac{\tilde{I}_{r,1}^3}{Z_1 Z_2} + \int_{\bm \ell}\frac{\tilde{I}_{r,1} \tilde{I}_{r,2}}{Z_1} \,,
\label{eq:example}
\end{align} 
where the sum over permutations of distinct $\tilde{I}_{r,n}$ is implicit;
for instance, $\tilde{I}_{r,1} \tilde{I}_{r,2} \equiv \tilde{I}_{r,1}(\bm \ell_1) \tilde{I}_{r,2}(\bm \ell_2)+\tilde{I}_{r,2}(\bm \ell_1) \tilde{I}_{r,1}(\bm \ell_2)$ while $\tilde{I}_{r,1}^3 \equiv \tilde{I}_{r,1}(\bm \ell_1) \tilde{I}_{r,1}(\bm \ell_2)\tilde{I}_{r,1}(\bm \ell_3)$.
As we can see, the classical part of the amplitude with this pole choice is directly the radial action, in contrast to Refs.~\cite{2PM,3PM, 3PMLong}.
We have explicitly verified \Eqs{eq:action}{eq:product} through $\mathcal{O}(G^4)$ by
comparing the amplitude calculation in EFT to the radial action from classical mechanics~\cite{4PMlong}. 
With some care, the relation \eqref{eq:action} can be established more generally to higher orders in $G$~\cite{4PMlong}.

%%%%%%%%%%%%%%%%%%%%%%%%%%%%%%%%%%%%
\Section{Constructing the Integrand.} The calculation of the amplitude through ${\cal  O}(G^4)$ begins with the construction of the classical limit of the
three-loop integrand. We use generalized unitarity, as described in Ref.~\cite{3PMLong}, which builds gravitational loop integrands directly from
on-shell tree amplitudes of scalars and gravitons.
The maximal-cut version~\cite{MaximalCutMethod} of generalized unitarity adopted here
organizes the cuts hierarchically.

As explained in Ref.~\cite{3PMLong}, potential contributions come from cuts with at least one matter line per loop and with no
gravitons beginning and ending on the same matter line. 
We drop contributions not of this type, such as self-energy loops and matter contact diagrams. The eight distinct
contributing generalized cuts are shown in Fig.~\ref{fig:cuts}, with
the others given by simple relabelings.

We use the $D$-dimensional tree-level BCJ double copy~\cite{BCJ} to obtain the
gravitational tree amplitudes from corresponding
gauge-theory ones.
The dilaton and antisymmetric tensor which naturally appear as
intermediate states are
straightforwardly eliminated by including graviton physical-state
projectors on the cut legs.  Conveniently, the reference light-cone momentum in
their definition cancels automatically, by organizing the tree
amplitudes so that they obey generalized Ward identities, following Ref.~\cite{Kosmopoulos:2020pcd}. A similar
strategy was used earlier for simpler one- and two-loop
calculations~\cite{3PM, Paolo2PM, 3PMLong}.

%%%%%%%%%%%%%% FIGURE %%%%%%%%%%%%%                             
\begin{figure}[t]
\begin{center}
\includegraphics[scale=.29]{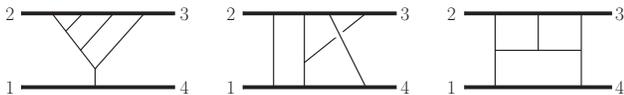}
\end{center}
\vskip -.4cm
\caption{Sample diagrams at ${\cal O}(G^4)$. From left to right: a contribution in the
 probe limit, a nonplanar diagram that
  contains iteration terms, and a diagram that contains contributions related to the tail
  effect.}
\vspace{-0.4cm}
\label{fig:diagrams}
\end{figure}
%%%%%%%%%%%%%%%%%%%%%%%%%%%%%%%%%%% 

The resulting integrand is then organized in terms of 
51 distinct cubic-vertex Feynman-like diagrams, of which
three are shown in Fig.~\ref{fig:diagrams}. All remaining diagrams are obtained
by relabeling the external momenta. The integrals are then further reduced to a
basis using integration by parts~\cite{IBP} implemented through
FIRE6~\cite{FIRE}, and graph symmetries. We isolate iteration
integrals, which cancel in the EFT matching prior to explicit integration.

Throughout, we work with rescaled variables $u_i = \bar{p}_i/|\bar{p}_i|$, following
Ref.~\cite{PRZ}, so that the integrals depend on the single variable $y = u_1 \cdot u_2$, enormously
simplifying the analysis. This also factors out the mass dependence, clearly exposing  the overlap between PM gravity and gravitational self-force~\cite{SpinMass, BuonannoEnergetics,
DamourSelfForce, SpinTutiFruiti}.  

\Section{Evaluating the Integrals.} Our strategy for integration combines the
nonrelativistic method of Refs.~\cite{2PM,3PM,3PMLong} and the method
of differential equations following Ref.~\cite{PRZ}.  The two methods
complement each other to systematically determine the analytic result
to all orders in velocity, and have been applied to multiple
examples~\cite{3PMFeynman,3PMworldline,BLRSZ,PMTidal1,UltraRelativisticLimit}.  

The nonrelativistic method evaluates
integrals as an expansion in powers of $v$, where the leading term serves as the boundary value for solving the differential equations. This also efficiently isolates the iterations in \Eq{eq:product}, which cancel directly in the EFT matching. The first step is to perform energy loop integration by
localizing to residues given by matter poles in the potential region. For a given integral ${\cal I}$ the
result of energy integration is 
\eq{ 
\int \prod_{i=1}^3 {d\omega_i \over 2\pi} {\cal
    I}(\omega_1, \omega_2, \omega_3) = \sum_{i} S_{i} \, {\rm Res }_i
  \, {\cal I}(\omega_1, \omega_2, \omega_3) \,,
}
{eq:residues} 
where the sum runs over triplets of matter poles on which the
residues are evaluated. The symmetry factors $S_{i}$ are determined
from the cuts in Fig.~\ref{fig:cuts}, building upon the
prescription in Ref.~\cite{3PMLong}. The remaining three-dimensional integrals are then expanded in $v$, and reduced to master integrals via integration by parts~\cite{IBP} using FIRE6~\cite{FIRE}. Iterations are identified by the pole structure in \Eq{eq:product}, and the final integrals are the same as in NRGR~\cite{NRGR}. 

Following Ref.~\cite{PRZ}, we use differential
equations to analytically solve integrals, or to obtain solutions
expanded in $v$ to very high orders. The boundary conditions for the
univariate differential equations are imposed in the leading PN expansion.  A
new feature at ${\cal O} (G^4)$ is the appearance of elliptic integrals, which
precludes a canonical basis~\cite{Henn}. 
This complicates the structure of the
differential equations. We solve part of the system exactly in terms of classical
polylogarithms up to weight three and complete elliptic integrals.  
The remaining contribution to the amplitude is determined by constructing an
ansatz as a linear combination of a subset of the known functions with rational
coefficients, whose parameters we fix using a series solution to the
differential equations up to ${\cal O}(v^{60})$.  In this way we obtain the
full velocity dependence, which can be checked, for instance, by obtaining a
series solution up to ${\cal O}(v^{400})$, finding perfect agreement. Details will be presented in
Ref.~\cite{4PMlong}.
%This complicates the structure of the
%differential equations. We solve part of the system in terms of classical
%polylogarithms up to weight three, and verify with the velocity expansion.  
%The elliptic contribution is determined by first obtaining complete solutions
%to the differential equations of a simple set of integrals in this class. Then
%an ansatz for the elliptic part of the amplitude is constructed as a linear
%combination of these with rational coefficients, whose parameters we fix using
%a series solution to the differential equations up to ${\cal O}(v^{60})$.  In
%this way we obtain the full velocity dependence, which can be checked, for
%instance, by obtaining a series solution up to ${\cal O}(v^{400})$. Details
%will be presented in Ref.~\cite{4PMlong}.

\Section{Amplitude.}
Performing this calculation, we obtain the following 4PM classical amplitude in the potential region: 
\begin{widetext}
  \begin{align}
    &{\cal M}_4(\bm q) = 
 G^4 M^7 \nu^2 |\bm q|\left(\frac{\bm q^2}{4^{1\over 3}\tilde\mu^2 }\right)^{-3\epsilon}\!\!\pi^2\left[{\cal M}_4^{\rm p} + \nu \left({ {\cal M}_4^{\rm t} \over \epsilon}   + {\cal M}_4^{\rm f} \right) \right]    
    + \int_{\bm \ell} \frac{\tilde{I}_{r,1}^4}{ Z_1  Z_2  Z_3 }
    + \int_{\bm \ell} \frac{\tilde{I}_{r,1}^2 \tilde{I}_{r,2} }{ Z_1  Z_2 }
    + \int_{\bm \ell} \frac{\tilde{I}_{r,1} \tilde{I}_{r,3}}{ Z_1} + \int_{\bm \ell} \frac{\tilde{I}_{r,2}^2}{ Z_1} \,,
    \nonumber \\[5pt]
    &{\cal M}_4^{\rm p} = 
%%%%% begin : M4pPaper
- \frac{35 \left(1-18 \sigma ^2+33 \sigma ^4\right)}{8 \left(\sigma ^2-1\right)} 
%%%%% end : M4pPaper
\,, 
\hskip 1.5 cm  {\cal M}_4^{\rm t} = 
%%%%% begin : M4tPaper
 h_1 + h_2 \log \left(\tfrac{\sigma +1}{2}\right) +  h_3\frac{  {\rm arccosh} (\sigma) }{\sqrt{\sigma^2-1}}
%%%%% end : M4tPaper
\,,
 \label{eq:amplitude} \\[5pt]
 &{\cal M}_4^{\rm f} =  
%%%%% begin : M4fPaper
   h_4
 + h_5 \log \left(\tfrac{\sigma +1}{2}\right) 
 + h_6  \frac{ {\rm arccosh}(\sigma )   }{\sqrt{\sigma^2-1}}
 + h_7 \log (\sigma ) 
 - h_2 \frac{2\pi^2}{3}  
 + h_8 \, { {\rm arccosh}^2(\sigma )  \over \sigma^2-1}
 + h_9 {}
 \left[\text{Li}_2\left(\tfrac{1-\sigma }{2}\right) + \frac12 \log ^2\left(\tfrac{\sigma +1}{2}\right)\right] 
 \nonumber \\
 &\quad 
 +  h_{10} 
 \left[\text{Li}_2\left(\tfrac{1-\sigma }{2}\right)  - \frac{\pi^2}{6}\right]  
 +  h_{11}
    \left[
       \text{Li}_2\left(\tfrac{1-\sigma }{1+\sigma }\right) 
      - \text{Li}_2\left(\tfrac{\sigma -1}{\sigma +1}\right) 
   + \frac{\pi^2}{3}\right] 
   +  h_{2} \frac{2 \sigma (2\sigma^2-3) }{(\sigma ^2-1)^{3/2}} 
    \left[ 
        \text{Li}_2\left(\sqrt{\tfrac{\sigma-1}{\sigma +1}}\right)
       -\text{Li}_2\left(-\sqrt{\tfrac{\sigma-1}{\sigma +1}}\right)\right]
	\nonumber \\
  &\quad
    + \frac{2h_3 }{\sqrt{\sigma^2-1}} {} \bigg[ \text{Li}_2\!\left(\!1\!-\!\sigma \!-\!\sqrt{\sigma ^2-1}\right) 
    \! - \!  \text{Li}_2\!\left(\!1\!-\!\sigma \!+\!\sqrt{\sigma ^2-1}\right) 
       +  5 \text{Li}_2\!\left(\!\sqrt{\tfrac{\sigma-1}{\sigma +1}}\right)
     \! - \!  5 \text{Li}_2\!\left(\!-\sqrt{\tfrac{\sigma-1}{\sigma +1}}\right) 
       +  2 \log \left(\tfrac{\sigma +1}{2}\right) {\rm arccosh} (\sigma)
  \bigg]
  \nonumber \\
&\quad  + h_{12} \K^2\left(\tfrac{\sigma -1}{\sigma +1}\right)+h_{13} \K \left(\tfrac{\sigma -1}{\sigma +1}\right) \E\left(\tfrac{\sigma -1}{\sigma +1}\right)+h_{14} \E^2\left(\tfrac{\sigma -1}{\sigma +1}\right) 
\nonumber  
%%%%% end : M4fPaper
\,,
\end{align}
\vspace{-15pt}
\end{widetext}
where $M=m_1+m_2$ is the total mass, $\nu=m_1m_2/M^2$ is the symmetric
mass ratio, $\tilde\mu^2 = 4 \pi \mu^2 e^{- \gamma_E}$ is the
renormalization scale in $\overline{\rm MS}$ scheme.  $\text{Li}_2$ is
the dilogarithm, and $\K$ and $\E$ are the complete elliptic integrals
of the first and second kind, respectively. The coefficient functions
$h_i$ are collected in Table~\ref{table:functions}.

%%%%%%%%%%%%%%%%%%%%%%%%%%%%%%%%%%%%%%%%%%%%
\begin{table}[tbh]
\setlength{\tabcolsep}{1pt} % Default value: 6pt
\renewcommand{\arraystretch}{3}
\begin{tabular}{|c|}
\hline
\scalebox{0.84}{\tabeq{10cm}{
 h_{1}  &= 
%%%%% begin : hPaper[1]
\frac{1151-3336 \sigma + 3148 \sigma^2 - 912 \sigma^3 + 339 \sigma^4 
      - 552 \sigma^5 + 210 \sigma^6} {12 (\sigma^2-1 ) }
%%%%% end : hPaper[1]
 \\
 h_{2}  &= 
%%%%% begin : hPaper[2]
\frac{1}{2} \left(5 - 76 \sigma  + 150 \sigma^2 - 60\sigma^3 - 35 \sigma^4 \right)
%%%%% end : hPaper[2]
 \\
h_3    &= 
%%%%% begin : hPaper[3]
\sigma \frac{\left(-3+2 \sigma^2\right)}{4(\sigma^2 - 1)}
 \left(11 - 30 \sigma^2 + 35 \sigma^4\right)
%%%%% end : hPaper[3]
\\
h_4  &=
%%%%% begin : hPaper[4]
\frac{1}{144
   \bigl(\sigma^2-1 )^2 \sigma^7} (-45+207 \sigma^2-1471 \sigma^4+13349 \sigma^6 \\
& \quad - 37566 \sigma^7+104753 \sigma^8 - 12312 \sigma^9-102759 \sigma^{10}-105498 \sigma^{11} \\
   &\quad + 134745 \sigma^{12} + 83844 \sigma^{13} - 101979 \sigma^{14} + 13644 \sigma^{15} + 10800 \sigma^{16} \bigr) 
%%%%% end : hPaper[4]
\\
h_5  &= 
%%%%% begin : hPaper[5]
\frac{1}{4 (\sigma^2 -1)} \bigl(
1759 - 4768 \sigma + 3407 \sigma^2 - 1316 \sigma^3 + 957 \sigma^4 \\
& \quad  - 672 \sigma^5 + 341 \sigma^6 + 100 \sigma^7 \bigr)
%%%%% end : hPaper[5]
\\
h_6   &= 
%%%%% begin : hPaper[6]
\frac{1}{24 (\sigma^2-1)^2}
\bigl(1237+7959 \sigma - 25183 \sigma^2 + 12915 \sigma^3 + 18102 \sigma^4 \\
& \quad - 12105 \sigma^5 - 9572 \sigma^6 + 2973 \sigma^7 + 5816 \sigma^8 - 2046 \sigma^9 \bigr)
%%%%% end : hPaper[6]
 \\
 h_7   &=
%%%%% begin : hPaper[7]
2 \sigma \frac{ \left(-852 - 283 \sigma^2 - 140 \sigma^4 + 75 \sigma^6\right)}{3 (\sigma^2 - 1)}
%%%%% end : hPaper[7]
\\
h_8 &=
%%%%% begin : hPaper[8]
\frac{\sigma}{8 (\sigma^2 -1)^2} \bigl(-304 - 99 \sigma + 672 \sigma^2 + 
  402 \sigma^3 - 192 \sigma^4 - 719 \sigma^5  \\ 
 & \quad  - 416 \sigma^6
  + 540 \sigma^7 + 240 \sigma^8 - 140 \sigma^9 \bigl)
%%%%% end : hPaper[8]
 \\
h_{9}   & =
%%%%% begin : hPaper[9]
\frac{1}{2} \left(52 - 532 \sigma + 351 \sigma^2 - 420 \sigma^3 + 30 \sigma^4 - 25 \sigma^6 \right)
%%%%% end : hPaper[9]
\\
h_{10}   &=
%%%%% begin : hPaper[10]
2 \left( 27 + 90 \sigma^2 + 35 \sigma^4 \right)
%%%%% end : hPaper[10]
\\
h_{11}   &= 
%%%%% begin : hPaper[11]
20 + 111 \sigma^2 + 30 \sigma^4 - 25 \sigma^6
%%%%% end : hPaper[11]
\\
h_{12}   &=
%%%%% begin : hPaper[12]
 \frac{834+2095 \sigma +1200 \sigma^2}{2 (\sigma^2-1 )}
%%%%% end : hPaper[12]
 \\
h_{13}   &=
%%%%% begin : hPaper[13]
  -\frac{1183 + 2929 \sigma + 2660 \sigma^2 + 1200 \sigma^3}{2 (\sigma^2-1 )} 
%%%%% end : hPaper[13]
\\
h_{14}   &= 
%%%%% begin : hPaper[14]
\frac{7 \left(169 + 380 \sigma^2\right)}{4 (\sigma-1 )} 
%%%%% end : hPaper[14]
   }}
   \\
\hline
\end{tabular}
\caption{Functions specifying the amplitude in \Eq{eq:amplitude}. }
\label{table:functions}
\end{table}
%%%%%%%%%%%%%%%%%%%%%%%%%%%%%%%%%%%%%%%%%%%%%%%%%%%%%%%%%%%%%%%%%%

We emphasize that \Eq{eq:amplitude} uses dimensional regularization with $D =
4-2\epsilon$ and that $\tilde{I}_{r,1}$, $\tilde{I}_{r,2}$ and
$\tilde{I}_{r,3}$ are expanded to the classical limit.  The tail effect
manifests as a ${1 /\epsilon}$ IR divergence in the classical term, due to the
overlap between potential and radiation contributions~\cite{Zerobin}. Including
the latter, which is not done here, would cancel this divergence and the
associated scheme dependence, replace ${\tilde \mu}$ with a physical scale, and
also add finite terms. Note that the scheme dependence starts at 4PN and enters
only through the coefficient functions $h_{4}$, $h_{5}$, and $h_{6}$.

The amplitude naturally exposes the simple dependence in the symmetric mass ratio
$\nu$, consistent with Ref.~\cite{DamourSelfForce}.  The leading term ${\cal M}^{\rm
  p}_4$ agrees with the result obtained using the Schwarzschild
solution~\cite{WS93,geodesic}. The next-to-leading terms ${\cal
  M}^{\rm t}_4$ and ${\cal M}^{\rm f}_4$ overlap with first-order
self-force~\cite{DamourSelfForce}.

As for the ${\cal O}(G^3)$ case, the ultrarelativistic limit of the
conservative result in \Eq{eq:amplitude} does not smoothly match onto
the massless case. The amplitude has a leading power
discontinuity of the form $\sim G^4 \bm p^8 |\bm q| (m_1 + m_2)/(m_1
m_2)$, consistent with dimensional analysis. One can expect this to
cancel with radiative effects~\cite{UltraRelativisticLimit,DamourRadiation}.

Given the relation in \Eq{eq:action}, it is straightforward to derive the radial action from the classical term in \Eq{eq:amplitude} via inverting \Eq{eq:FT_J}
\begin{align}
I_{r,4}(J) =&-\frac{G^4 M^7 \nu^2\pi  {\bm p}^2}{8 E J^3} \left(\frac{4 \tilde{\mu}^2 e^{2\gamma_E} J^2}{\bm p^2} \right)^{4\epsilon}   \nn 
\\
&\times \left[{\cal M}_4^{\rm p} + \nu \left({ {\cal M}_4^{\rm t} \over \epsilon}   + {\cal M}_4^{\rm f}-14{\cal M}_4^{\rm t} \right) \right] \,,
\label{eq:radial_4}
\end{align}
which inherits the simple mass dependence from the amplitude~\cite{Bini:2020wpo,DamourSelfForce}.
Moreover, we have checked that $I_{r,1}$, $I_{r,2}$ and $I_{r,3}$ obtained from the iteration terms in \Eq{eq:amplitude} are consistent with the known results \cite{B2B}.

The scattering angle is then given by $\chi=-\partial I_r/\partial J$. We compare to the ${\cal O}(G^4)$ scattering angle obtained from potential contributions to the Hamiltonian up to 5PN, given in Eqs.~(21)--(26) of~\cite{Blumlein4PN} and Eq.~(5) of~\cite{Blumlein5PN}.\footnote{Since this work first appeared the authors have extended their calculation to 6PN \cite{Blumlein6PN}, finding agreement with our results.} We find agreement including terms that depend on conventional dimensional regularization, which first enter at 4PN. We also compare the regularization-scheme-independent $\pi^3$ terms with the 6PN result in Eq.~(8.4) of Ref.~\cite{Bini:2020nsb}, and find agreement. Here we have only focused on potential region contributions to the dynamics. A comparison with the full 4PN Hamiltonian \cite{4PNDJS,4PNrest} and remaining 5PN and 6PN parts of the scattering angle \cite{Bini:2020nsb,Bini:2020hmy,Bini:2020wpo} would require radiation contributions, which we have not pursued here.

As discussed in Refs.~\cite{Bini:2017wfr, BFLS, Bini:2020hmy}, the 
tail term ${\cal M}^{\rm t}_4$ is related to the energy loss $\Delta E$ from radiation emission, and we thus identify
\eq{
  \Delta E = \frac{G^3 M^7 \nu^3 \pi  \bm p^2}{4 E^2 J^3} {\cal M}^{\rm t}_4\,,
}{}  
in the COM frame. We have compared this to the direct calculation of the energy loss in Ref.~\cite{HPRZ} using the formalism of Ref.~\cite{Kosower:2018adc}, finding agreement. Additional checks of $\Delta E$ are discussed in Ref.~\cite{HPRZ}. We can also obtain other observables for bound orbits via analytic continuation~\cite{B2B}; details will be presented elsewhere~\cite{4PMlong}.

\Section{Hamiltonian.}
Following the approach in Ref.~\cite{2PM}, we can construct the two-body Hamiltonian in isotropic gauge
\eq{
H^{\rm iso} = E_1+ E_2 + \sum_{n=1}^\infty {G^n  (r^2 \tilde{\mu}^2 e^{2\gamma_E})^{n\epsilon}\over r^n} c_n(\bm p^2)\,,
}{eq:H}
where $r$ is the distance between bodies, and $E_i = \sqrt{\bm p^2 + m_i^2}$ are the energies of the incoming particles.
The $\mathcal{O}(\epsilon)$ corrections are relevant when $c_n(\bm p^2)$ has divergence, which first occurs at ${\cal O}(G^4)$ in isotropic gauge.
The coefficients $c_n(\bm p^2)$ are determined by matching as in Refs.~\cite{2PM,3PM,3PMLong}, but using the new pole choice in \Eq{eq:product}.
Upon accounting for this we find,
\begin{widetext}
\eq{
c_4 &=
%%%%% begin : c4Paper
 {M^7 \nu^2 \over 4\xi E^2} \left[{\cal M}_4^{\rm p} + \nu {} \left({ {\cal M}_4^{\rm t} \over \epsilon}   + {\cal M}_4^{\rm f}-10{\cal M}_4^{\rm t} \right) \right] 
+ {\cal D}^3 \left[ {E^3 \xi^3 \over 3} c_1^4 \right] +{\cal D}^2 \left[ \left(\frac{E^3 \xi ^3}{\bm p^2}+\frac{E \xi {} (3 \xi-1 )}{2} \right) c_1^4 -2 E^2 \xi^2 c_1^2 c_2 \right] 
\\
&\quad 
+\left( {\cal D} + {1 \over \bm p^2} \right) \bigg[  E \xi {} (2c_1 c_3 + c_2^2) + \left( \frac{4\xi-1}{4E}+\frac{2 E^3 \xi ^3}{\bm p^4}+\frac{E \xi (3 \xi-1 )}{\bm p^2} \right)c_1^4 + \left((1-3 \xi )-\frac{4 E^2 \xi ^2}{\bm p^2}\right) c_1^2 c_2\bigg]
%%%%% end : c4Paper
\,,
}{}
\end{widetext}
where $\xi = E_1E_2/E^2$, and ${\cal D} = {d \over d \bm p^2}$ denotes
differentiation with respect to $\bm p^2$.  The lower-order coefficients
$c_1$, $c_2$, and $c_3$ can be found in Eq.~(10) of
Ref.~\cite{3PM}.  The final explicit result for $c_4$ is included
in the ancillary file~\cite{AttachedFile}.  Note that the iteration terms in
\Eq{eq:amplitude} cancel in this matching, providing another
nontrivial check.  

\Section{Conclusions.} 
In this letter we applied and extended amplitudes and EFT based methods to
determine the potential contribution to binary dynamics at ${\cal O}(G^4)$,
offering a first look at the PM tail effect.  Radiation contributions need to
be added to completely determine the ${\cal O}(G^4)$ dynamics.  It would be
important to investigate the application of our methods to this problem.  It
would also be interesting to study closed-orbit observables at ${\cal O}(G^4)$,
as was done for ${\cal O}(G^3)$~\cite{BuonannoEnergetics}, and via analytic
continuation~\cite{B2B}.  The interplay with gravitational
self-force~\cite{self_force, SpinMass, SelfForceReview, DamourSelfForce,
SpinTutiFruiti,Barack:2019agd}, the structure of long-distance logarithms at
higher orders via renormalization group techniques~\cite{BFLS,Galley, RGrefs},
and the complete tail contribution in the PM framework also deserve further
study. 
 
Aside from the obvious application to gravitational-wave physics, our
calculation elucidates emerging structures and identifies new tools. The
amplitude-action relation in \Eq{eq:action} greatly clarifies the link between
scattering amplitudes and classical mechanics. It is natural to expect that
this structure holds more generally.  Obvious extensions within the PM
framework include spin~\cite{PMSpin,SpinMass, BLRSZ},
tidal~\cite{PMTidal1,PMTidal,geodesic}, and
radiation~\cite{PMRadiation,SoftRadiation} effects. Most excitingly, the
methods applied here are not close to being exhausted.

\Section{Acknowledgments.}
We thank Samuel Abreu, Johannes Bl\"umlein, Alessandra Buonanno, Clifford
Cheung, Thibault Damour, Lance Dixon, Enrico Herrmann, Andr\'{e}s Luna, Rafael
Porto, Ira Rothstein, Jan Steinhoff, Gabriele Veneziano and Justin Vines for
helpful discussions.  Z.B.~is supported by the U.S. Department of Energy (DOE)
under Award Number DE-SC0009937.  J.P.-M.~is supported by the U.S.\ Department
of Energy (DOE) under award number~DE-SC0011632.  R.R.~is supported by the U.S.
Department of Energy (DOE) under grant no.~DE-SC0013699.  M.S.R.’s work is
funded by the German Research Foundation (DFG) within the Research Training
Group GRK 2044.  C.-H.S. is supported by the U.S.\ Department of Energy (DOE)
under award number~DE-SC0009919.  M.P.S. is supported by the David Saxon
Presidential Term Chair.  M.Z.'s work is supported by the U.K.\ Royal Society
through Grant URF{\textbackslash}R1{\textbackslash}20109.  We thank the Mani L.
Bhaumik Institute for support.
%%%%%%%%%%%%%%%%%%%%


\begin{thebibliography}{99}

\bibitem{LIGO}
B.~P.~Abbott {\it et al.} [LIGO Scientific and Virgo Collaborations],
%``Observation of gravitational waves from a binary black hole merger,''
Phys.\ Rev.\ Lett.\  {\bf 116}, no. 6, 061102 (2016)
%doi:10.1103/PhysRevLett.116.061102
[arXiv:1602.03837 [gr-qc]];
%%CITATION = doi:10.1103/PhysRevLett.116.061102;%% 
% 
 B.~P.~Abbott {\it et al.} [LIGO Scientific and Virgo Collaborations],
%``GW170817: Observation of gravitational waves from a binary neutron star inspiral,''
Phys.\ Rev.\ Lett.\  {\bf 119}, no. 16, 161101 (2017)
%doi:10.1103/PhysRevLett.119.161101
[arXiv:1710.05832 [gr-qc]].
%%CITATION = doi:10.1103/PhysRevLett.119.161101;%%

\bibitem{NewDetectors}
M.~Punturo {\it et al.,} ``The Einstein Telescope: A third-generation gravitational wave observatory,''
 Class. Quant. Grav. 27 (2010) 194002;
%
P.~Amaro-Seoane \textit{et al.} [LISA],
%``Laser Interferometer Space Antenna,''
[arXiv:1702.00786 [astro-ph.IM]];
%844 citations counted in INSPIRE as of 05 Oct 2020
%
D.~Reitze \textit{et al.}
%``Cosmic Explorer: The U.S. Contribution to Gravitational-Wave Astronomy beyond LIGO,''
Bull. Am. Astron. Soc. \textbf{51}, 035
[arXiv:1907.04833 [astro-ph.IM]].
%120 citations counted in INSPIRE as of 30 Dec 2020

\bibitem{2PM}
C.~Cheung, I.~Z.~Rothstein and M.~P.~Solon,
%``From Scattering Amplitudes to Classical Potentials in the Post-Minkowskian Expansion,''
Phys. Rev. Lett. \textbf{121}, no.25, 251101 (2018)
%doi:10.1103/PhysRevLett.121.251101
[arXiv:1808.02489 [hep-th]].
%121 citations counted in INSPIRE as of 27 Dec 2020

\bibitem{3PM}
Z.~Bern, C.~Cheung, R.~Roiban, C.~H.~Shen, M.~P.~Solon and M.~Zeng,
%``Scattering Amplitudes and the Conservative Hamiltonian for Binary Systems at Third Post-Minkowskian Order,''
Phys. Rev. Lett. \textbf{122}, no.20, 201603 (2019)
%doi:10.1103/PhysRevLett.122.201603
[arXiv:1901.04424 [hep-th]].
%139 citations counted in INSPIRE as of 27 Dec 2020

\bibitem{3PMLong}
Z.~Bern, C.~Cheung, R.~Roiban, C.~H.~Shen, M.~P.~Solon and M.~Zeng,
%``Black Hole Binary Dynamics from the Double Copy and Effective Theory,''
JHEP \textbf{10}, 206 (2019)
%doi:10.1007/JHEP10(2019)206
[arXiv:1908.01493 [hep-th]].
%112 citations counted in INSPIRE as of 27 Dec 2020

\bibitem{ScatteringToGrav}
Y.~Iwasaki,
%``Quantum theory of gravitation vs. classical theory---fourth-order potential,''
Prog.\ Theor.\ Phys.\  {\bf 46}, 1587 (1971);
%doi:10.1143/PTP.46.1587;\\
%%CITATION = doi:10.1143/PTP.46.1587;%%
%                                                                                                                                 
Y.~Iwasaki,
%``Fourth-order gravitational potential based on quantum field theory,''
Lett.\ Nuovo Cim.\  {\bf 1S2}, 783 (1971)
[Lett.\ Nuovo Cim.\  {\bf 1}, 783 (1971)];
%doi:10.1007/BF02770190;\\
%%CITATION = doi:10.1007/BF02770190;%% 
%                                                                                                                                 
S.~N.~Gupta and S.~F.~Radford,
%``Improved gravitational coupling of scalar fields,''
Phys.\ Rev.\ D {\bf 19}, 1065 (1979).
%doi:10.1103/PhysRevD.19.1065.
%%CITATION = doi:10.1103/PhysRevD.19.1065;%%

\bibitem{RecentAmpToPotential}
J.~F.~Donoghue,
%``General relativity as an effective field theory: The leading quantum corrections,''
Phys. Rev. D \textbf{50}, 3874-3888 (1994)
%doi:10.1103/PhysRevD.50.3874
[arXiv:gr-qc/9405057 [gr-qc]];
%
N.~E.~J.~Bjerrum-Bohr, J.~F.~Donoghue and B.~R.~Holstein,
%``Quantum gravitational corrections to the nonrelativistic scattering potential of two masses,''
Phys. Rev. D \textbf{67}, 084033 (2003)
[erratum: Phys. Rev. D \textbf{71}, 069903 (2005)]
%doi:10.1103/PhysRevD.71.069903
[arXiv:hep-th/0211072 [hep-th]];
%289 citations counted in INSPIRE as of 14 Jan 2021
%
N.~E.~J.~Bjerrum-Bohr, J.~F.~Donoghue and P.~Vanhove,
%``On-shell techniques and universal results in quantum gravity,''
JHEP {\bf 1402}, 111 (2014)
%doi:10.1007/JHEP02(2014)111
[arXiv:1309.0804 [hep-th]].
%%CITATION = doi:10.1007/JHEP02(2014)111;%%

\bibitem{NeillandRothstein}
D.~Neill and I.~Z.~Rothstein,
%``Classical Space-Times from the S Matrix,''
Nucl. Phys. B \textbf{877}, 177-189 (2013)
%doi:10.1016/j.nuclphysb.2013.09.007
[arXiv:1304.7263 [hep-th]].
%98 citations counted in INSPIRE as of 30 Dec 2020

\bibitem{NRGR}
W.~D.~Goldberger and I.~Z.~Rothstein,
%``An effective field theory of gravity for extended objects,''
Phys.\ Rev.\ D {\bf 73}, 104029 (2006)
%doi:10.1103/PhysRevD.73.104029
[hep-th/0409156].
%%CITATION = doi:10.1103/PhysRevD.73.104029;%%

\bibitem{GeneralizedUnitarity}
Z.~Bern, L.~J.~Dixon, D.~C.~Dunbar and D.~A.~Kosower,
%``One loop $n$-point gauge-theory amplitudes, unitarity and collinear limits,''
Nucl.\ Phys.\ B {\bf 425}, 217 (1994)
%doi:10.1016/0550-3213(94)90179-1
[hep-ph/9403226];
%%CITATION = doi:10.1016/0550-3213(94)90179-1;%%
%
Z.~Bern, L.~J.~Dixon, D.~C.~Dunbar and D.~A.~Kosower,
%``Fusing gauge-theory tree amplitudes into loop amplitudes,''
Nucl.\ Phys.\ B {\bf 435}, 59 (1995)
%doi:10.1016/0550-3213(94)00488-Z
[hep-ph/9409265];
%%CITATION = doi:10.1016/0550-3213(94)00488-Z;%%
%
Z.~Bern and A.~G.~Morgan,
%``Massive loop amplitudes from unitarity,''
Nucl.\ Phys.\ B {\bf 467}, 479 (1996)
%doi:10.1016/0550-3213(96)00078-8
[hep-ph/9511336];
%%CITATION = doi:10.1016/0550-3213(96)00078-8;%%
%
Z.~Bern, L.~J.~Dixon and D.~A.~Kosower,
%``One loop amplitudes for $e^+ e^-$ to four partons,''
Nucl.\ Phys.\ B {\bf 513}, 3 (1998)
%doi:10.1016/S0550-3213(97)00703-7
[hep-ph/9708239];
%%CITATION = doi:10.1016/S0550-3213(97)00703-7;%%
%
R.~Britto, F.~Cachazo and B.~Feng,
%``Generalized unitarity and one-loop amplitudes in $N=4$ super-Yang-Mills,''
Nucl.\ Phys.\ B {\bf 725}, 275 (2005)
%doi:10.1016/j.nuclphysb.2005.07.014
[hep-th/0412103].
%%CITATION = doi:10.1016/j.nuclphysb.2005.07.014;%%

\bibitem{KLT}
H.~Kawai, D.~C.~Lewellen and S.~H.~H.~Tye,
%``A relation between tree amplitudes of closed and open strings,''
Nucl.\ Phys.\ B {\bf 269}, 1 (1986);
%doi:10.1016/0550-3213(86)90362-7;
%%CITATION = doi:10.1016/0550-3213(86)90362-7;%%
%
Z.~Bern, L.~J.~Dixon, M.~Perelstein and J.~S.~Rozowsky,
%``Multileg one loop gravity amplitudes from gauge theory,''
Nucl.\ Phys.\ B {\bf 546}, 423 (1999)
%doi:10.1016/S0550-3213(99)00029-2
[hep-th/9811140].
%%CITATION = doi:10.1016/S0550-3213(99)00029-2;%%

\bibitem{BCJ}
Z.~Bern, J.~J.~M.~Carrasco and H.~Johansson,
%``New relations for gauge-theory amplitudes,''
Phys.\ Rev.\ D {\bf 78}, 085011 (2008)
%doi:10.1103/PhysRevD.78.085011
[arXiv:0805.3993 [hep-ph]];
%%CITATION = doi:10.1103/PhysRevD.78.085011;%%                                                             
%                                                                                                          
Z.~Bern, J.~J.~M.~Carrasco and H.~Johansson,
%``Perturbative quantum gravity as a double copy of gauge theory,''
Phys.\ Rev.\ Lett.\  {\bf 105}, 061602 (2010)
%doi:10.1103/PhysRevLett.105.061602                                                                        
[arXiv:1004.0476 [hep-th]];
%%CITATION = doi:10.1103/PhysRevLett.105.061602;%%
%                                                                                                          
Z.~Bern, J.~J.~M.~Carrasco, W.~M.~Chen, H.~Johansson, R.~Roiban and M.~Zeng,
%``Five-loop four-point integrand of $N=8$ supergravity as a generalized double copy,''
Phys. Rev. D \textbf{96}, no.12, 126012 (2017)
%doi:10.1103/PhysRevD.96.126012
[arXiv:1708.06807 [hep-th]];
%
Z.~Bern, J.~J.~Carrasco, W.~M.~Chen, A.~Edison, H.~Johansson, J.~Parra-Martinez, R.~Roiban and M.~Zeng,
%``Ultraviolet Properties of $\mathcal N = 8$ Supergravity at Five Loops,''
Phys. Rev. D \textbf{98}, no.8, 086021 (2018)
%doi:10.1103/PhysRevD.98.086021
[arXiv:1804.09311 [hep-th]];
%
Z.~Bern, J.~J.~Carrasco, M.~Chiodaroli, H.~Johansson and R.~Roiban,
%``The duality between color and kinematics and its applications,''
arXiv:1909.01358 [hep-th].
%%CITATION = ARXIV:1909.01358;%%

\bibitem{IBP}
K.G.~Chetyrkin and F.V.~Tkachov,
%``Integration by parts: the algorithm to calculate beta functions in 4 loops,''
Nucl.\ Phys.\ B {\bf 192}, 159 (1981);
%%CITATION = NUPHA,B192,159;%% 
%
S.~Laporta,
%``High precision calculation of multiloop Feynman integrals by difference equations,''
Int. J. Mod. Phys. A \textbf{15}, 5087-5159 (2000)
%doi:10.1016/S0217-751X(00)00215-7
[arXiv:hep-ph/0102033 [hep-ph]].
%930 citations counted in INSPIRE as of 04 Oct 2020 

\bibitem{DEs}
A.~V.~Kotikov,
%``Differential equations method: New technique for massive Feynman diagrams calculation,''
Phys. Lett. B \textbf{254}, 158-164 (1991);
%doi:10.1016/0370-2693(91)90413-K
%590 citations counted in INSPIRE as of 15 Oct 2020
%
Z.~Bern, L.~J.~Dixon and D.~A.~Kosower,
%``Dimensionally regulated pentagon integrals,''
Nucl. Phys. B \textbf{412}, 751-816 (1994)
%doi:10.1016/0550-3213(94)90398-0
[arXiv:hep-ph/9306240 [hep-ph]];
%487 citations counted in INSPIRE as of 05 Oct 2020
%
E.~Remiddi,
%``Differential equations for Feynman graph amplitudes,''
Nuovo Cim. A \textbf{110}, 1435-1452 (1997)
[arXiv:hep-th/9711188 [hep-th]];
%474 citations counted in INSPIRE as of 15 Oct 2020
%
T.~Gehrmann and E.~Remiddi,
%``Differential equations for two loop four point functions,''
Nucl. Phys. B \textbf{580}, 485-518 (2000)
%doi:10.1016/S0550-3213(00)00223-6
[arXiv:hep-ph/9912329 [hep-ph]].
%642 citations counted in INSPIRE as of 05 Oct 2020

\bibitem{FIRE}
A.~V.~Smirnov,
%``Algorithm FIRE -- Feynman Integral REduction,''
JHEP \textbf{10}, 107 (2008)
%doi:10.1088/1126-6708/2008/10/107
[arXiv:0807.3243 [hep-ph]];
%427 citations counted in INSPIRE as of 30 Sep 2020
%
A.~V.~Smirnov and F.~S.~Chuharev,
%``FIRE6: Feynman Integral REduction with Modular Arithmetic,''
%doi:10.1016/j.cpc.2019.106877
[arXiv:1901.07808 [hep-ph]].
%64 citations counted in INSPIRE as of 30 Sep 2020 

\bibitem{6PNBlumlein}
J.~Bl\"umlein, A.~Maier, P.~Marquard and G.~Sch\"afer,
%``Testing binary dynamics in gravity at the sixth post-Newtonian level,''
arXiv:2003.07145 [gr-qc];
%%CITATION = ARXIV:2003.07145;%%                                                                           
\bibitem{3PMFeynman}
C.~Cheung and M.~P.~Solon,
%``Classical gravitational scattering at ${\cal O}(G^3)$ from Feynman diagrams,''
arXiv:2003.08351 [hep-th];
%%CITATION = ARXIV:2003.08351;%%
\bibitem{3PMworldline}
G.~K\"alin, Z.~Liu and R.~A.~Porto,
%``Conservative Dynamics of Binary Systems to Third Post-Minkowskian Order from the Effective Field Theory Approach,''
[arXiv:2007.04977 [hep-th]].

\bibitem{Bini:2020wpo}
D.~Bini, T.~Damour and A.~Geralico,
%``Binary dynamics at the fifth and fifth-and-a-half post-Newtonian orders,''
Phys. Rev. D \textbf{102}, no.2, 024062 (2020)
%doi:10.1103/PhysRevD.102.024062
[arXiv:2003.11891 [gr-qc]].
%33 citations counted in INSPIRE as of 14 Jan 2021

\bibitem{Bini:2020nsb}
D.~Bini, T.~Damour and A.~Geralico,
%``Sixth post-Newtonian local-in-time dynamics of binary systems,''
Phys. Rev. D \textbf{102}, no.2, 024061 (2020)
%doi:10.1103/PhysRevD.102.024061
[arXiv:2004.05407 [gr-qc]].
%26 citations counted in INSPIRE as of 14 Jan 2021

\bibitem{EOB}
A.~Buonanno and T.~Damour,
%``Effective one-body approach to general relativistic two-body dynamics,''
Phys.\ Rev.\ D {\bf 59}, 084006 (1999)
%doi:10.1103/PhysRevD.59.084006
[gr-qc/9811091];
%%CITATION = doi:10.1103/PhysRevD.59.084006;%% 
%
A.~Buonanno and T.~Damour,
%``Transition from inspiral to plunge in binary black hole coalescences,''
Phys.\ Rev.\ D {\bf 62}, 064015 (2000)
%doi:10.1103/PhysRevD.62.064015
[gr-qc/0001013].
%%CITATION = doi:10.1103/PhysRevD.62.064015;%%

\bibitem{NR}
F.~Pretorius,
%``Evolution of binary black hole spacetimes,''
Phys.\ Rev.\ Lett.\  {\bf 95}, 121101 (2005)
%doi:10.1103/PhysRevLett.95.121101
[gr-qc/0507014];
%%CITATION = doi:10.1103/PhysRevLett.95.121101;%%
%
M.~Campanelli, C.~O.~Lousto, P.~Marronetti and Y.~Zlochower,
%``Accurate evolutions of orbiting black-hole binaries without excision,''
Phys.\ Rev.\ Lett.\  {\bf 96}, 111101 (2006)
%doi:10.1103/PhysRevLett.96.111101
[gr-qc/0511048];
%%CITATION = doi:10.1103/PhysRevLett.96.111101;%%
%
J.~G.~Baker, J.~Centrella, D.~I.~Choi, M.~Koppitz and J.~van Meter,
%``Gravitational wave extraction from an inspiraling configuration of merging black holes,''
Phys.\ Rev.\ Lett.\  {\bf 96}, 111102 (2006)
%doi:10.1103/PhysRevLett.96.111102
[gr-qc/0511103].
%%CITATION = doi:10.1103/PhysRevLett.96.111102;%%

\bibitem{self_force} 
Y.~Mino, M.~Sasaki and T.~Tanaka,
%``Gravitational radiation reaction to a particle motion,''
Phys.\ Rev.\ D {\bf 55}, 3457 (1997)
%doi:10.1103/PhysRevD.55.3457
[gr-qc/9606018];
%%CITATION = doi:10.1103/PhysRevD.55.3457;%%
%
T.~C.~Quinn and R.~M.~Wald,
%``An Axiomatic approach to electromagnetic and gravitational radiation reaction of particles in curved space-time,''
Phys.\ Rev.\ D {\bf 56}, 3381 (1997)
%doi:10.1103/PhysRevD.56.3381
[gr-qc/9610053].
%%CITATION = doi:10.1103/PhysRevD.56.3381;%%

\bibitem{SelfForceReview}
E.~Poisson, A.~Pound and I.~Vega,
%``The Motion of point particles in curved spacetime,''
Living Rev. Rel. \textbf{14}, 7 (2011)
%doi:10.12942/lrr-2011-7
[arXiv:1102.0529 [gr-qc]];
%402 citations counted in INSPIRE as of 09 Jan 2021
%
L.~Barack and A.~Pound,
%``Self-force and radiation reaction in general relativity,''
Rept. Prog. Phys. \textbf{82}, no.1, 016904 (2019)
%doi:10.1088/1361-6633/aae552
[arXiv:1805.10385 [gr-qc]].

\bibitem{PN}
J.~Droste,
%``The field of $n$ moving centres in Einstein's theory of gravitation,''
Proc.\ Acad.\ Sci.\ Amst.\ 19:447–455 (1916);
%                                  
H. A. Lorentz and J. Droste, 
%``De beweging van een stelsel lichamen onder de theorie van Einstein I,II''
Koninklijke Akademie Van Wetenschappen te Amsterdam 26 392, 649 (1917).
English translation in ``Lorentz Collected papers,'' P. Zeeman and A. D. Fokker editors, Vol 5, 330 (1934-1939),
The Hague: Nijhof;
%
A.~Einstein, L.~Infeld and B.~Hoffmann,
%``The Gravitational equations and the problem of motion,''
Annals Math.\  {\bf 39}, 65 (1938);
%doi:10.2307/1968714;
%%CITATION = doi:10.2307/1968714;%%
%
% 2PN             
T.~Ohta, H.~Okamura, T.~Kimura and K.~Hiida,
%``Physically acceptable solution of einstein's equation for many-body system,''
Prog.\ Theor.\ Phys.\  {\bf 50}, 492 (1973);
%doi:10.1143/PTP.50.492;
%%CITATION = doi:10.1143/PTP.50.492;%%            
%
T.~Damour and N.~Deruelle,
%``Radiation Reaction and Angular Momentum Loss in Small Angle Gravitational Scattering,''
Phys.\ Lett.\ A {\bf 87}, 81 (1981);
%doi:10.1016/0375-9601(81)90567-3
T.~Damour;
%``Probl\`eme des deux corps et freinage de rayonnement en relativit\'e g\'en\'erale,''
C.R.\ Acad.\ Sc.\ Paris, S\'erie II, {\bf 294}, pp 1355-1357 (1982);
%
% 3PN             
P.~Jaranowski and G.~Sch\"afer,
%``Third post-Newtonian higher order ADM Hamilton dynamics for two-body point mass systems,''
Phys.\ Rev.\ D {\bf 57}, 7274 (1998)
Erratum: [Phys.\ Rev.\ D {\bf 63}, 029902 (2000)]
%doi:10.1103/PhysRevD.57.7274, 10.1103/PhysRevD.63.029902            
[gr-qc/9712075];
%%CITATION = doi:10.1103/PhysRevD.57.7274, 10.1103/PhysRevD.63.029902;%%              
%
T.~Damour, P.~Jaranowski and G.~Sch\"afer,
%``Dynamical invariants for general relativistic two-body systems at the third         
%post-Newtonian approximation,''
Phys.\ Rev.\ D {\bf 62}, 044024 (2000)
%doi:10.1103/PhysRevD.62.044024    
[gr-qc/9912092];
%%CITATION = doi:10.1103/PhysRevD.62.044024;%%      
%
L.~Blanchet and G.~Faye,
%``Equations of motion of point particle binaries at the third post-Newtonian order,''
Phys.\ Lett.\ A {\bf 271}, 58 (2000)
%doi:10.1016/S0375-9601(00)00360-1 
[gr-qc/0004009];
%%CITATION = doi:10.1016/S0375-9601(00)00360-1;%%   
%
L.~Blanchet and G.~Faye,
%``General relativistic dynamics of compact binaries at the third postNewtonian order,''
Phys. Rev. D \textbf{63}, 062005 (2001)
%doi:10.1103/PhysRevD.63.062005
[arXiv:gr-qc/0007051 [gr-qc]];
%163 citations counted in INSPIRE as of 20 Jan 2021
%
V.~C.~de Andrade, L.~Blanchet and G.~Faye,
%``Third postNewtonian dynamics of compact binaries: Noetherian conserved quantities and equivalence between the harmonic coordinate and ADM Hamiltonian formalisms,''
Class. Quant. Grav. \textbf{18}, 753-778 (2001)
%doi:10.1088/0264-9381/18/5/301
[arXiv:gr-qc/0011063 [gr-qc]];
%145 citations counted in INSPIRE as of 20 Jan 2021
%
T.~Damour, P.~Jaranowski and G.~Sch\"afer,
%``Dimensional regularization of the gravitational interaction of point masses,''
Phys.\ Lett.\ B {\bf 513}, 147 (2001)
%doi:10.1016/S0370-2693(01)00642-6 
[gr-qc/0105038];
%%CITATION = doi:10.1016/S0370-2693(01)00642-6;%%   
%
L.~Blanchet and B.~R.~Iyer,
%``Third postNewtonian dynamics of compact binaries: Equations of motion in the center-of-mass frame,''
Class. Quant. Grav. \textbf{20}, 755 (2003)
%doi:10.1088/0264-9381/20/4/309
[arXiv:gr-qc/0209089 [gr-qc]];
%111 citations counted in INSPIRE as of 21 Jan 2021
%
L.~Blanchet, T.~Damour and G.~Esposito-Farese,
%``Dimensional regularization of the third postNewtonian dynamics of point particles in harmonic coordinates,''
Phys. Rev. D \textbf{69}, 124007 (2004)
%doi:10.1103/PhysRevD.69.124007
[arXiv:gr-qc/0311052 [gr-qc]];
%178 citations counted in INSPIRE as of 21 Jan 2021
% 4PN            

\bibitem{4PNDJS}
T.~Damour, P.~Jaranowski and G.~Sch\"afer,
%``Nonlocal-in-time action for the fourth post-Newtonian conservative dynamics of two-body systems,''
Phys.\ Rev.\ D {\bf 89}, no. 6, 064058 (2014)
%doi:10.1103/PhysRevD.89.064058    
[arXiv:1401.4548 [gr-qc]];
%%CITATION = doi:10.1103/PhysRevD.89.064058;%%      
%
P.~Jaranowski and G.~Sch\"afer,
%``Derivation of local-in-time fourth post-Newtonian ADM Hamiltonian for spinless compact binaries,''
Phys.\ Rev.\ D {\bf 92}, no. 12, 124043 (2015)
%doi:10.1103/PhysRevD.92.124043    
[arXiv:1508.01016 [gr-qc]];
%%CITATION = doi:10.1103/PhysRevD.92.124043;%%      

\bibitem{4PNrest}
L.~Bernard, L.~Blanchet, A.~Boh\'e, G.~Faye and S.~Marsat,
%``Fokker action of nonspinning compact binaries at the fourth post-Newtonian approximation,''
Phys. Rev. D \textbf{93}, no.8, 084037 (2016)
%doi:10.1103/PhysRevD.93.084037    
[arXiv:1512.02876 [gr-qc]];
%78 citations counted in INSPIRE as of 15 Oct 2020  
%    
L.~Bernard, L.~Blanchet, A.~Boh\'e, G.~Faye and S.~Marsat,
%``Energy and periastron advance of compact binaries on circular orbits at the fourth post-Newtonian order,''
Phys. Rev. D \textbf{95}, no.4, 044026 (2017)
%doi:10.1103/PhysRevD.95.044026    
[arXiv:1610.07934 [gr-qc]];
%66 citations counted in INSPIRE as of 15 Oct 2020  
%
L.~Bernard, L.~Blanchet, A.~Boh\'e, G.~Faye and S.~Marsat,
%``Dimensional regularization of the IR divergences in the Fokker action of point-particle binaries at the fourth post-Newtonian order,''
Phys. Rev. D \textbf{96}, no.10, 104043 (2017)
%doi:10.1103/PhysRevD.96.104043    
[arXiv:1706.08480 [gr-qc]];
%
T.~Marchand, L.~Bernard, L.~Blanchet and G.~Faye,
%``Ambiguity-Free Completion of the Equations of Motion of Compact Binary Systems at the Fourth Post-Newtonian Order,''
Phys. Rev. D \textbf{97}, no.4, 044023 (2018)
%doi:10.1103/PhysRevD.97.044023    
[arXiv:1707.09289 [gr-qc]];
%
L.~Bernard, L.~Blanchet, G.~Faye and T.~Marchand,
%``Center-of-Mass Equations of Motion and Conserved Integrals of Compact Binary Systems at the Fourth Post-Newtonian Order,''
Phys. Rev. D \textbf{97}, no.4, 044037 (2018)
%doi:10.1103/PhysRevD.97.044037    
[arXiv:1711.00283 [gr-qc]].
%41 citations counted in INSPIRE as of 15 Oct 2020    

\bibitem{Bini:2017wfr}
D.~Bini and T.~Damour,
%``Gravitational scattering of two black holes at the fourth post-Newtonian approximation,''
Phys. Rev. D \textbf{96}, no.6, 064021 (2017)
%doi:10.1103/PhysRevD.96.064021
[arXiv:1706.06877 [gr-qc]].
%32 citations counted in INSPIRE as of 16 Jan 2021

\bibitem{PM}
B. Bertotti,
%``On  gravitational  motion'',
Nuovo  Cimento {\bf 4}:898-906  (1956)
%doi:10.1007/BF02746175;
%
R. P. Kerr,
%``The Lorentz-covariant approximation method in general relativity'',
I. Nuovo Cimento {\bf 13}:469-491 (1959);
%doi:10.1007/BF02732767;
%
B.~Bertotti, J.~F.~Pleba\'nski,
%``Theory of gravitational perturbations in the fast motion approximation'',
Ann, Phys, {\bf 11}:169-200 (1960);
%doi:10.1016/0003-4916(60)90132-9;
%
M.~Portilla,
% ``Momentum and angular momentum of two gravitating particles,''
J.\ Phys.\ A {\bf 12}, 1075 (1979);
%doi:10.1088/0305-4470/12/7/025;
%%CITATION = doi:10.1088/0305-4470/12/7/025;%%
%
K.~Westpfahl and M.~Goller,
%``Gravitational scattering of two relativistic particles in postlinear approximation,''
Lett.\ Nuovo Cim.\  {\bf 26}, 573 (1979)
%doi:10.1007/BF02817047;
%%CITATION = doi:10.1007/BF02817047;%%
%  
M.~Portilla,
%``Scattering of two gravitating particles: classical approach,''
J.\ Phys.\ A {\bf 13}, 3677 (1980)
%doi:10.1088/0305-4470/13/12/017;
%%CITATION = doi:10.1088/0305-4470/13/12/017;%%
%
L.~Bel, T.~Damour, N.~Deruelle, J.~Ibanez and J.~Martin,
%``Poincar\'e-invariant gravitational field and equations of motion of two pointlike objects: The postlinear approximation of general relativity,''
Gen.\ Rel.\ Grav.\  {\bf 13}, 963 (1981);
% doi:10.1007/BF00756073;
%%CITATION = doi:10.1007/BF00756073;%%
%
K. Westpfahl,
%``High-speed scattering of charged and uncharged particles in general relativity'',
Fortschr. Phys., {\bf 33}, 417 (1985);
%
T.~Ledvinka, G.~Schaefer and J.~Bicak,
%``Relativistic closed-form Hamiltonian for many-body gravitating systems in the post-Minkowskian approximation,''
Phys.\ Rev.\ Lett.\  {\bf 100}, 251101 (2008)
% doi:10.1103/PhysRevLett.100.251101
[arXiv:0807.0214 [gr-qc]].
%%CITATION = doi:10.1103/PhysRevLett.100.251101;%%

\bibitem{DamourPM}
T.~Damour,
%``Gravitational scattering, post-Minkowskian approximation and effective one-body theory,''
Phys.\ Rev.\ D {\bf 94}, no. 10, 104015 (2016)
%doi:10.1103/PhysRevD.94.104015
[arXiv:1609.00354 [gr-qc]];
%%CITATION = doi:10.1103/PhysRevD.94.104015;%%                                                                                    
%
T.~Damour,
%``High-energy gravitational scattering and the general relativistic two-body problem,''
Phys.\ Rev.\ D {\bf 97}, no. 4, 044038 (2018)
%  doi:10.1103/PhysRevD.97.044038
[arXiv:1710.10599 [gr-qc]].
%%CITATION = doi:10.1103/PhysRevD.97.044038;%%                                                                                  

\bibitem{NRGRrecent}
B.~Kol and M.~Smolkin,
%``Classical effective field theory and caged black holes,''
Phys. Rev. D \textbf{77}, 064033 (2008)
%doi:10.1103/PhysRevD.77.064033
[arXiv:0712.2822 [hep-th]];
%61 citations counted in INSPIRE as of 19 Apr 2020
%
B.~Kol and M.~Smolkin,
%``Non-Relativistic gravitation: from Newton to Einstein and back,''
Class.\ Quant.\ Grav.\  {\bf 25}, 145011 (2008)
%doi:10.1088/0264-9381/25/14/145011
[arXiv:0712.4116 [hep-th]];
%%CITATION = %doi:10.1088/0264-9381/25/14/145011;%%
%71 citations counted in INSPIRE as of 25 Mar 2020
%
J.~B.~Gilmore and A.~Ross,
%``Effective field theory calculation of second post-Newtonian binary dynamics,''
Phys.\ Rev.\ D {\bf 78}, 124021 (2008)
%doi:10.1103/PhysRevD.78.124021
[arXiv:0810.1328 [gr-qc]];
%%CITATION = %doi:10.1103/PhysRevD.78.124021;%%
%71 citations counted in INSPIRE as of 25 Mar 2020
%
B.~Kol, M.~Levi and M.~Smolkin,
%``Comparing space+time decompositions in the post-Newtonian limit,''
Class. Quant. Grav. \textbf{28}, 145021 (2011)
%doi:10.1088/0264-9381/28/14/145021
[arXiv:1011.6024 [gr-qc]];
%
S.~Foffa and R.~Sturani,
%``Effective field theory calculation of conservative binary dynamics at third post-Newtonian order,''
Phys.\ Rev.\ D {\bf 84}, 044031 (2011)
%doi:10.1103/PhysRevD.84.044031
[arXiv:1104.1122 [gr-qc]];
%%CITATION = doi:10.1103/PhysRevD.84.044031;%%
%70 citations counted in INSPIRE as of 25 Mar 2020
%
S.~Foffa and R.~Sturani,
%``Dynamics of the gravitational two-body problem at fourth post-Newtonian order and at quadratic order in the Newton constant,''
Phys. Rev. D \textbf{87}, no.6, 064011 (2013)
%doi:10.1103/PhysRevD.87.064011
[arXiv:1206.7087 [gr-qc]];
%77 citations counted in INSPIRE as of 14 Jan 2021
%
S.~Foffa, P.~Mastrolia, R.~Sturani and C.~Sturm,
%``Effective field theory approach to the gravitational two-body dynamics, at fourth post-Newtonian order and quintic in the Newton constant,''
Phys.\ Rev.\ D {\bf 95}, no. 10, 104009 (2017)
%doi:10.1103/PhysRevD.95.104009
[arXiv:1612.00482 [gr-qc]];
%%CITATION = doi:10.1103/PhysRevD.95.104009;%%
%41 citations counted in INSPIRE as of 25 Mar 2020
%
S.~Foffa, P.~Mastrolia, R.~Sturani, C.~Sturm and W.~J.~Torres Bobadilla,
%``Static two-body potential at fifth post-Newtonian order,''
Phys.\ Rev.\ Lett.\  {\bf 122}, no. 24, 241605 (2019)
%doi:10.1103/PhysRevLett.122.241605
[arXiv:1902.10571 [gr-qc]];
%%CITATION = doi:10.1103/PhysRevLett.122.241605;%%
%
J.~Bl\"umlein, A.~Maier and P.~Marquard,
%``Five-Loop static contribution to the gravitational interaction potential of two point masses,''
Phys.\ Lett.\ B {\bf 800}, 135100 (2020)
%doi:10.1016/j.physletb.2019.135100
[arXiv:1902.11180 [gr-qc]];
%%CITATION = doi:10.1016/j.physletb.2019.135100;%%
%
S.~Foffa and R.~Sturani,
%``Conservative dynamics of binary systems to fourth Post-Newtonian order in the EFT approach I: Regularized Lagrangian,''
Phys.\ Rev.\ D {\bf 100}, no. 2, 024047 (2019)
%doi:10.1103/PhysRevD.100.024047
[arXiv:1903.05113 [gr-qc]];
%%CITATION = doi:10.1103/PhysRevD.100.024047;%%
%18 citations counted in INSPIRE as of 25 Mar 2020
%
S.~Foffa, R.~A.~Porto, I.~Rothstein and R.~Sturani,
%``Conservative dynamics of binary systems to fourth Post-Newtonian order in the EFT approach II: Renormalized Lagrangian,''
Phys.\ Rev.\ D {\bf 100}, no. 2, 024048 (2019)
%doi:10.1103/PhysRevD.100.024048
[arXiv:1903.05118 [gr-qc]];
%%CITATION = doi:10.1103/PhysRevD.100.024048;%%
%
S.~Foffa, R.~Sturani and W.~J.~Torres Bobadilla,
%``Efficient resummation of high post-Newtonian contributions to the binding energy,''
[arXiv:2010.13730 [gr-qc]].
%0 citations counted in INSPIRE as of 15 Jan 2021

\bibitem{NBI}
N.~E.~J.~Bjerrum-Bohr, A.~Cristofoli and P.~H.~Damgaard,
%``Post-Minkowskian Scattering Angle in Einstein Gravity,''
JHEP \textbf{08}, 038 (2020)
%doi:10.1007/JHEP08(2020)038
[arXiv:1910.09366 [hep-th]].
%43 citations counted in INSPIRE as of 08 Jan 2021

\bibitem{PMModernList}
F.~Cachazo and A.~Guevara,
%``Leading singularities and classical gravitational scattering,''
JHEP {\bf 2002}, 181 (2020)
%doi:10.1007/JHEP02(2020)181
[arXiv:1705.10262 [hep-th]];
%%CITATION = doi:10.1007/JHEP02(2020)181;%%                                                                                       
%
N.~E.~J.~Bjerrum-Bohr, P.~H.~Damgaard, G.~Festuccia, L.~Plant\'e and P.~Vanhove,
%``General relativity from scattering amplitudes,''
Phys.\ Rev.\ Lett.\  {\bf 121}, no. 17, 171601 (2018)
%doi:10.1103/PhysRevLett.121.171601
[arXiv:1806.04920 [hep-th]];
%%CITATION = doi:10.1103/PhysRevLett.121.171601;%%                                                                               
%
A.~Cristofoli, N.~E.~J.~Bjerrum-Bohr, P.~H.~Damgaard and P.~Vanhove,
%``Post-Minkowskian Hamiltonians in general relativity,''
Phys. Rev. D \textbf{100}, no.8, 084040 (2019)
%doi:10.1103/PhysRevD.100.084040
[arXiv:1906.01579 [hep-th]];
%51 citations counted in INSPIRE as of 10 Jan 2021
%
P.~H.~Damgaard, K.~Haddad and A.~Helset,
%``Heavy Black Hole Effective Theory,‘’
JHEP \textbf{11}, 070 (2019)
%doi:10.1007/JHEP11(2019)070
[arXiv:1908.10308 [hep-ph]];
%
A.~Cristofoli, P.~H.~Damgaard, P.~Di Vecchia and C.~Heissenberg,
%``Second-order Post-Minkowskian scattering in arbitrary dimensions,‘’
JHEP \textbf{07}, 122 (2020)
%doi:10.1007/JHEP07(2020)122
[arXiv:2003.10274 [hep-th]];
%                                                                                                          
G.~K\"alin and R.~A.~Porto,
%``Post-Minkowskian Effective Field Theory for Conservative Binary Dynamics,''
[arXiv:2006.01184 [hep-th]];
%
G.~Mogull, J.~Plefka and J.~Steinhoff,
%``Classical black hole scattering from a worldline quantum field theory,''
[arXiv:2010.02865 [hep-th]].
%11 citations counted in INSPIRE as of 15 Jan 2021

\bibitem{Kosower:2018adc}
D.~A.~Kosower, B.~Maybee and D.~O'Connell,
%``Amplitudes, Observables, and Classical Scattering,''
JHEP \textbf{02}, 137 (2019)
%doi:10.1007/JHEP02(2019)137
[arXiv:1811.10950 [hep-th]].

\bibitem{Paolo2PM}
A.~Koemans Collado, P.~Di Vecchia and R.~Russo,
%``Revisiting the second post-Minkowskian eikonal and the dynamics of binary black holes,''
Phys. Rev. D \textbf{100}, no.6, 066028 (2019)
%doi:10.1103/PhysRevD.100.066028
[arXiv:1904.02667 [hep-th]].
%50 citations counted in INSPIRE as of 10 Jan 2021

\bibitem{B2B}
G.~K\"alin and R.~A.~Porto,
%``From Boundary Data to Bound States,''
JHEP \textbf{01}, 072 (2020)
%doi:10.1007/JHEP01(2020)072
[arXiv:1910.03008 [hep-th]];
%54 citations counted in INSPIRE as of 08 Jan 2021
%
G.~K\"alin and R.~A.~Porto,
%``From boundary data to bound states. Part II. Scattering angle to dynamical invariants (with twist),''
JHEP \textbf{02}, 120 (2020)
%doi:10.1007/JHEP02(2020)120
[arXiv:1911.09130 [hep-th]].
%34 citations counted in INSPIRE as of 08 Jan 2021

\bibitem{DamourSelfForce}
T.~Damour,
%``Classical and quantum scattering in post-Minkowskian gravity,''
Phys. Rev. D \textbf{102}, no.2, 024060 (2020)
%doi:10.1103/PhysRevD.102.024060
[arXiv:1912.02139 [gr-qc]];
%36 citations counted in INSPIRE as of 02 Jan 2021
%
D.~Bini, T.~Damour and A.~Geralico,
%``Novel approach to binary dynamics: application to the fifth post-Newtonian level,''
Phys. Rev. Lett. \textbf{123}, no.23, 231104 (2019)
%doi:10.1103/PhysRevLett.123.231104
[arXiv:1909.02375 [gr-qc]].
%37 citations counted in INSPIRE as of 30 Dec 2020

\bibitem{SpinTutiFruiti}
A.~Antonelli, C.~Kavanagh, M.~Khalil, J.~Steinhoff and J.~Vines,
%``Gravitational spin-orbit coupling through third-subleading post-Newtonian order: from first-order self-force to arbitrary mass ratios,''
Phys. Rev. Lett. \textbf{125}, no.1, 011103 (2020)
%doi:10.1103/PhysRevLett.125.011103
[arXiv:2003.11391 [gr-qc]];
%13 citations counted in INSPIRE as of 09 Jan 2021
%
A.~Antonelli, C.~Kavanagh, M.~Khalil, J.~Steinhoff and J.~Vines,
%``Gravitational spin-orbit and aligned spin$_1$-spin$_2$ couplings through third-subleading post-Newtonian orders,''
Phys. Rev. D \textbf{102}, 124024 (2020)
%doi:10.1103/PhysRevD.102.124024
[arXiv:2010.02018 [gr-qc]].

\bibitem{Bini:2020hmy}
D.~Bini, T.~Damour and A.~Geralico,
%``Sixth post-Newtonian nonlocal-in-time dynamics of binary systems,''
Phys. Rev. D \textbf{102}, no.8, 084047 (2020)
%doi:10.1103/PhysRevD.102.084047
[arXiv:2007.11239 [gr-qc]].
%8 citations counted in INSPIRE as of 28 Dec 2020

\bibitem{BFLS}
L.~Blanchet, S.~Foffa, F.~Larrouturou and R.~Sturani,
%``Logarithmic tail contributions to the energy function of circular compact binaries,''
Phys. Rev. D \textbf{101}, no.8, 084045 (2020)
%doi:10.1103/PhysRevD.101.084045
[arXiv:1912.12359 [gr-qc]].
  
\bibitem{TailEffect}
W.~Bonnor, Philos. Trans. R. Soc. London, Ser. A 251, 233 (1959);
%                                                                                                                    
W.~Bonnor and M.~Rotenberg, Proc. R. Soc. London, Ser. A 289, 247 (1966);
%
K.~S.~Thorne,
%``Multipole expansions of gravitational radiation,''
Rev.\ Mod.\ Phys.\  {\bf 52}, 299 (1980);
%doi:10.1103/RevModPhys.52.299;
%%CITATION = doi:10.1103/RevModPhys.52.299;%%
%
L.~Blanchet and T.~Damour,
%``Tail transported temporal correlations in the dynamics of a gravitating system,''
Phys.\ Rev.\ D {\bf 37}, 1410 (1988);
%doi:10.1103/PhysRevD.37.1410;
%%CITATION = doi:10.1103/PhysRevD.37.1410;%%
%
L.~Blanchet and T.~Damour,
%``Hereditary effects in gravitational radiation,''
Phys.\ Rev.\ D {\bf 46}, 4304 (1992);
%doi:10.1103/PhysRevD.46.4304;
%%CITATION = doi:10.1103/PhysRevD.46.4304;%%
%
L.~Blanchet and G.~Schaefer,
%``Gravitational wave tails and binary star systems,''
Class. Quant. Grav. \textbf{10}, 2699-2721 (1993).
%doi:10.1088/0264-9381/10/12/026
%137 citations counted in INSPIRE as of 15 Jan 2021

\bibitem{Galley}
S.~Foffa and R.~Sturani,
%``Tail terms in gravitational radiation reaction via effective field theory,''
Phys. Rev. D \textbf{87} (2013) no.4, 044056
%doi:10.1103/PhysRevD.87.044056
[arXiv:1111.5488 [gr-qc]];
%49 citations counted in INSPIRE as of 26 Feb 2021
C.~R.~Galley, A.~K.~Leibovich, R.~A.~Porto and A.~Ross,
%``Tail effect in gravitational radiation reaction: Time nonlocality and renormalization group evolution,''
Phys.\ Rev.\ D {\bf 93}, 124010 (2016)
%doi:10.1103/PhysRevD.93.124010
[arXiv:1511.07379 [gr-qc]];
%%CITATION = doi:10.1103/PhysRevD.93.124010;%%
S.~Foffa and R.~Sturani,
%``Hereditary terms at next-to-leading order in two-body gravitational dynamics,''
Phys. Rev. D \textbf{101}, no.6, 064033 (2020)
%doi:10.1103/PhysRevD.101.064033
[arXiv:1907.02869 [gr-qc]].
%16 citations counted in INSPIRE as of 15 Jan 2021

\bibitem{Beneke:1997zp}
M.~Beneke and V.~A.~Smirnov,
%``Asymptotic expansion of Feynman integrals near threshold,''
Nucl. Phys. B \textbf{522}, 321-344 (1998)
%doi:10.1016/S0550-3213(98)00138-2
[arXiv:hep-ph/9711391 [hep-ph]].
%667 citations counted in INSPIRE as of 11 Jan 2021

\bibitem{Zerobin}
A.~V.~Manohar and I.~W.~Stewart,
%``The Zero-Bin and Mode Factorization in Quantum Field Theory,''
Phys. Rev. D \textbf{76}, 074002 (2007)
%doi:10.1103/PhysRevD.76.074002
[arXiv:hep-ph/0605001 [hep-ph]];
%287 citations counted in INSPIRE as of 14 Jan 2021
%
R.~A.~Porto and I.~Z.~Rothstein,
%``Apparent ambiguities in the post-Newtonian expansion for binary systems,''
Phys.\ Rev.\ D {\bf 96}, no. 2, 024062 (2017)
%doi:10.1103/PhysRevD.96.024062
[arXiv:1703.06433 [gr-qc]].
%%CITATION = doi:10.1103/PhysRevD.96.024062;%%

\bibitem{DamourRadial}
T.~Damour and G.~Schaefer,
%``Higher Order Relativistic Periastron Advances and Binary Pulsars,''
Nuovo Cim. B \textbf{101}, 127 (1988);
%doi:10.1007/BF02828697
%210 citations counted in INSPIRE as of 08 Jan 2021
%
T.~Damour, P.~Jaranowski and G.~Schaefer,
%``Dynamical invariants for general relativistic two-body systems at the third postNewtonian approximation,''
Phys. Rev. D \textbf{62}, 044024 (2000)
%doi:10.1103/PhysRevD.62.044024
[arXiv:gr-qc/9912092 [gr-qc]].
%119 citations counted in INSPIRE as of 08 Jan 2021

\bibitem{Landshoff:1989ig}
P.~V.~Landshoff and J.~C.~Polkinghorne,
%``Iterations of regge cuts,''
Phys. Rev. \textbf{181}, 1989-1995 (1969).
%doi:10.1103/PhysRev.181.1989
%25 citations counted in INSPIRE as of 11 Jan 2021

\bibitem{PRZ}
J.~Parra-Martinez, M.~S.~Ruf and M.~Zeng,
%``Extremal black hole scattering at $\mathcal{O}(G^3)$: graviton dominance, eikonal exponentiation, and differential equations,''
JHEP \textbf{11}, 023 (2020)
%doi:10.1007/JHEP11(2020)023
[arXiv:2005.04236 [hep-th]].

\bibitem{EikonalPapers}
R.~J.~Glauber,
in ``Lectures in theoretical physics'', ed.~by W.~E. Brittin and L.~G.~Dunham,
Interscience Publishers, Inc., New York, Volume I, page 315, (1959);
%
D.~Amati, M.~Ciafaloni and G.~Veneziano,
%``Higher order gravitational deflection and soft bremsstrahlung in planckian energy superstring collisions,''
Nucl.\ Phys.\ B {\bf 347}, 550 (1990);
%doi:10.1016/0550-3213(90)90375-N;
%%CITATION = doi:10.1016/0550-3213(90)90375-N;%%
%217 citations counted in INSPIRE as of 20 Feb 2020
%
R.~Saotome and R.~Akhoury,
%``Relationship Between Gravity and Gauge Scattering in the High Energy Limit,''
JHEP \textbf{01}, 123 (2013)
%doi:10.1007/JHEP01(2013)123
[arXiv:1210.8111 [hep-th]];
%72 citations counted in INSPIRE as of 14 Jan 2021
%
R.~Akhoury, R.~Saotome and G.~Sterman,
%``High energy scattering in perturbative quantum gravity at next to leading power,''
arXiv:1308.5204 [hep-th];
%%CITATION = ARXIV:1308.5204;%%
%
P.~Di Vecchia, A.~Luna, S.~G.~Naculich, R.~Russo, G.~Veneziano and C.~D.~White,
%``A tale of two exponentiations in ${\cal N}=8$ supergravity,''
Phys. Lett. B \textbf{798}, 134927 (2019)
%doi:10.1016/j.physletb.2019.134927
[arXiv:1908.05603 [hep-th]];
%22 citations counted in INSPIRE as of 12 Jan 2021
%
P.~Di Vecchia, S.~G.~Naculich, R.~Russo, G.~Veneziano and C.~D.~White,
%``A tale of two exponentiations in $ \mathcal{N} $ = 8 supergravity at subleading level,''
JHEP \textbf{03}, 173 (2020)
%doi:10.1007/JHEP03(2020)173
[arXiv:1911.11716 [hep-th]];
%22 citations counted in INSPIRE as of 02 Jan 2021
%
Z.~Bern, H.~Ita, J.~Parra-Martinez and M.~S.~Ruf,
%``Universality in the classical limit of massless gravitational scattering,''
Phys. Rev. Lett. \textbf{125}, no.3, 031601 (2020)
%doi:10.1103/PhysRevLett.125.031601
[arXiv:2002.02459 [hep-th]].

\bibitem{4PMlong}
Z.~Bern, J.~Parra-Martinez, R.~Roiban, M.~S.~Ruf C.~H.~Shen, M.~P.~Solon and M.~Zeng,
to appear.

\bibitem{MaximalCutMethod}
Z.~Bern, J.~J.~M.~Carrasco, H.~Johansson and D.~A.~Kosower,
%``Maximally supersymmetric planar Yang-Mills amplitudes at five loops,''
Phys.\ Rev.\ D {\bf 76}, 125020 (2007)
%doi:10.1103/PhysRevD.76.125020
[arXiv:0705.1864 [hep-th]].
%%CITATION = doi:10.1103/PhysRevD.76.125020;%%

\bibitem{Kosmopoulos:2020pcd}
D.~Kosmopoulos,
%``Simplifying $D$-Dimensional Physical-State Sums in Gauge Theory and Gravity,''
[arXiv:2009.00141 [hep-th]].

\bibitem{SpinMass}
J.~Vines, J.~Steinhoff and A.~Buonanno,
%``Spinning-black-hole scattering and the test-black-hole limit at second post-Minkowskian order,''
Phys. Rev. D \textbf{99}, no.6, 064054 (2019)
%doi:10.1103/PhysRevD.99.064054
[arXiv:1812.00956 [gr-qc]].
%42 citations counted in INSPIRE as of 06 Jan 2021

\bibitem{BuonannoEnergetics}
A.~Antonelli, A.~Buonanno, J.~Steinhoff, M.~van de Meent and J.~Vines,
%``Energetics of two-body Hamiltonians in post-Minkowskian gravity,''
Phys. Rev. D \textbf{99}, no.10, 104004 (2019)
%doi:10.1103/PhysRevD.99.104004
[arXiv:1901.07102 [gr-qc]].
%62 citations counted in INSPIRE as of 08 Jan 2021

\bibitem{BLRSZ}
Z.~Bern, A.~Luna, R.~Roiban, C.~H.~Shen and M.~Zeng,
%``Spinning Black Hole Binary Dynamics, Scattering Amplitudes and Effective Field Theory,''
[arXiv:2005.03071 [hep-th]].

\bibitem{PMTidal1}
C.~Cheung and M.~P.~Solon,
%``Tidal Effects in the Post-Minkowskian Expansion,''
Phys. Rev. Lett. \textbf{125}, no.19, 191601 (2020)
%doi:10.1103/PhysRevLett.125.191601
[arXiv:2006.06665 [hep-th]];
%
G.~K\"alin, Z.~Liu and R.~A.~Porto,
%``Conservative Tidal Effects in Compact Binary Systems to Next-to-Leading Post-Minkowskian Order,''
Phys. Rev. D \textbf{102}, 124025 (2020)
%doi:10.1103/PhysRevD.102.124025
[arXiv:2008.06047 [hep-th]];
%
%
Z.~Bern, J.~Parra-Martinez, R.~Roiban, E.~Sawyer and C.~H.~Shen,
%``Leading Nonlinear Tidal Effects and Scattering Amplitudes,''
[arXiv:2010.08559 [hep-th]].

\bibitem{UltraRelativisticLimit}
P.~Di Vecchia, C.~Heissenberg, R.~Russo and G.~Veneziano,
%``Universality of ultra-relativistic gravitational scattering,''
Phys. Lett. B \textbf{811}, 135924 (2020)
%doi:10.1016/j.physletb.2020.135924
[arXiv:2008.12743 [hep-th]];
%13 citations counted in INSPIRE as of 30 Dec 2020
%
P.~Di Vecchia, C.~Heissenberg, R.~Russo and G.~Veneziano,
%``Radiation Reaction from Soft Theorems,''
[arXiv:2101.05772 [hep-th]].
%0 citations counted in INSPIRE as of 15 Jan 2021

\bibitem{Henn} 
J.~M.~Henn,
%``Multiloop integrals in dimensional regularization made simple,''
Phys. Rev. Lett. \textbf{110}, 251601 (2013)
%doi:10.1103/PhysRevLett.110.251601
[arXiv:1304.1806 [hep-th]];
%485 citations counted in INSPIRE as of 10 Jan 2021
%
J.~M.~Henn, A.~V.~Smirnov and V.~A.~Smirnov,
%``Evaluating single-scale and/or non-planar diagrams by differential equations,''
JHEP \textbf{03}, 088 (2014)
%doi:10.1007/JHEP03(2014)088
[arXiv:1312.2588 [hep-th]].
%93 citations counted in INSPIRE as of 10 Jan 2021

\bibitem{WS93} 
N.\ Wex and G.\ Sch\"afer, 
%``Innermost stable orbits for coalescing binary systems of compact objects-a remark''
Class.\ Quantum Grav.\ {\bf 10}, 2729 (1993).
%{http://stacks.iop.org/0264-9381/10/i=12/a=028}

\bibitem{geodesic}
C.~Cheung, N.~Shah and M.~P.~Solon,
%``Mining the Geodesic Equation for Scattering Data,''
[arXiv:2010.08568 [hep-th]].
%5 citations counted in INSPIRE as of 28 Dec 2020

\bibitem{DamourRadiation}
T.~Damour,
%``Radiative contribution to classical gravitational scattering at the third order in $G$,''
Phys. Rev. D \textbf{102}, no.12, 124008 (2020)
%doi:10.1103/PhysRevD.102.124008
[arXiv:2010.01641 [gr-qc]].
%8 citations counted in INSPIRE as of 08 Jan 2021

\bibitem{Blumlein4PN}
J.~Bl\"umlein, A.~Maier, P.~Marquard and G.~Sch\"afer,
%``Fourth post-Newtonian Hamiltonian dynamics of two-body systems from an effective field theory approach,''
Nucl. Phys. B \textbf{955}, 115041 (2020)
%doi:10.1016/j.nuclphysb.2020.115041
[arXiv:2003.01692 [gr-qc]].
%20 citations counted in INSPIRE as of 28 Dec 2020

\bibitem{Blumlein5PN}
J.~Bl\"umlein, A.~Maier, P.~Marquard and G.~Sch\"afer,
%``The fifth-order post-Newtonian Hamiltonian dynamics of two-body systems from an effective field theory approach: potential contributions,''
[arXiv:2010.13672 [gr-qc]].
%2 citations counted in INSPIRE as of 28 Dec 2020

\bibitem{Blumlein6PN}
J.~Bl\"umlein, A.~Maier, P.~Marquard and G.~Sch\"afer,
%``The 6th Post-Newtonian Potential Terms at $O(G_N^4)$,''
[arXiv:2101.08630 [gr-qc]].
%4 citations counted in INSPIRE as of 05 Mar 2021

\bibitem{HPRZ}
E.~Herrmann, J.~Parra-Martinez, M.~S.~Ruf and M.~Zeng,
%``Gravitational Bremsstrahlung from Reverse Unitarity,''
Phys. Rev. Lett. \textbf{126}, no.20, 201602 (2021)
doi:10.1103/PhysRevLett.126.201602
[arXiv:2101.07255 [hep-th]].

\bibitem{AttachedFile}
See the ancillary files of this manuscript.

\bibitem{Barack:2019agd}
L.~Barack, M.~Colleoni, T.~Damour, S.~Isoyama and N.~Sago,
%``Self-force effects on the marginally bound zoom-whirl orbit in Schwarzschild spacetime,''
Phys. Rev. D \textbf{100}, no.12, 124015 (2019)
%doi:10.1103/PhysRevD.100.124015
[arXiv:1909.06103 [gr-qc]].
%3 citations counted in INSPIRE as of 13 Jan 2021

\bibitem{RGrefs}
W.~D.~Goldberger and A.~Ross,
%``Gravitational radiative corrections from effective field theory,''
Phys. Rev. D \textbf{81}, 124015 (2010)
%doi:10.1103/PhysRevD.81.124015
[arXiv:0912.4254 [gr-qc]];
%
W.~D.~Goldberger, A.~Ross and I.~Z.~Rothstein,
%``Black hole mass dynamics and renormalization group evolution,''
Phys. Rev. D \textbf{89}, no.12, 124033 (2014)
%doi:10.1103/PhysRevD.89.124033
[arXiv:1211.6095 [hep-th]].


\bibitem{PMSpin}
J.~Vines,
%``Scattering of two spinning black holes in post-Minkowskian gravity, to all orders in spin, and effective-one-body mappings,''
Class.\ Quant.\ Grav.\  {\bf 35}, no. 8, 084002 (2018)
%doi:10.1088/1361-6382/aaa3a8
[arXiv:1709.06016 [gr-qc]].
%%CITATION = doi:10.1088/1361-6382/aaa3a8;%%
%
D.~Bini and T.~Damour,
%``Gravitational spin-orbit coupling in binary systems, post-Minkowskian approximation and effective one-body theory,''
Phys.\ Rev.\ D {\bf 96}, no. 10, 104038 (2017)
%doi:10.1103/PhysRevD.96.104038
[arXiv:1709.00590 [gr-qc]];
%%CITATION = doi:10.1103/PhysRevD.96.104038;%%
%
D.~Bini and T.~Damour,
%``Gravitational spin-orbit coupling in binary systems at the second post-Minkowskian approximation,''
Phys. Rev. D \textbf{98}, no.4, 044036 (2018)
%doi:10.1103/PhysRevD.98.044036
[arXiv:1805.10809 [gr-qc]];
%19 citations counted in INSPIRE as of 02 May 2020
%
A.~Guevara, A.~Ochirov and J.~Vines,
%``Scattering of Spinning Black Holes from Exponentiated Soft Factors,''
JHEP {\bf 1909}, 056 (2019)
%doi:10.1007/JHEP09(2019)056
[arXiv:1812.06895 [hep-th]];
%
A.~Guevara, A.~Ochirov and J.~Vines,
%``Black-hole scattering with general spin directions from minimal-coupling amplitudes,''
Phys.\ Rev.\ D {\bf 100}, no. 10, 104024 (2019)
%doi:10.1103/PhysRevD.100.104024
[arXiv:1906.10071 [hep-th]];
%
M.~Z.~Chung, Y.~T.~Huang, J.~W.~Kim and S.~Lee,
%``The simplest massive S-matrix: from minimal coupling to Black Holes,''
JHEP \textbf{04}, 156 (2019)
%doi:10.1007/JHEP04(2019)156
[arXiv:1812.08752 [hep-th]];
%78 citations counted in INSPIRE as of 14 Jan 2021
%
B.~Maybee, D.~O'Connell and J.~Vines,
%``Observables and amplitudes for spinning particles and black holes,''
JHEP \textbf{12}, 156 (2019)
%doi:10.1007/JHEP12(2019)156
[arXiv:1906.09260 [hep-th]];
%
%75 citations counted in INSPIRE as of 15 Jan 2021
H.~Johansson and A.~Ochirov,
%``Double copy for massive quantum particles with spin,''
JHEP \textbf{09}, 040 (2019)
%doi:10.1007/JHEP09(2019)040
[arXiv:1906.12292 [hep-th]];
%52 citations counted in INSPIRE as of 15 Jan 2021
%
Y.~F.~Bautista and A.~Guevara,
%``On the Double Copy for Spinning Matter,''
[arXiv:1908.11349 [hep-th]];
%31 citations counted in INSPIRE as of 15 Jan 2021
%
N.~Siemonsen and J.~Vines,
%``Test black holes, scattering amplitudes and perturbations of Kerr spacetime,''
Phys. Rev. D \textbf{101}, no.6, 064066 (2020)
%doi:10.1103/PhysRevD.101.064066
[arXiv:1909.07361 [gr-qc]];
%
R.~Aoude, K.~Haddad and A.~Helset,
%``On-shell heavy particle effective theories,‘’
JHEP \textbf{05}, 051 (2020)
%doi:10.1007/JHEP05(2020)051
[arXiv:2001.09164 [hep-th]];
%
M.~Z.~Chung, Y.~t.~Huang, J.~W.~Kim and S.~Lee,
%``Complete Hamiltonian for spinning binary systems at first post-Minkowskian order,''
arXiv:2003.06600 [hep-th];
%
K.~Haddad and A.~Helset,
%``The double copy for heavy particles,''
Phys. Rev. Lett. \textbf{125}, 181603 (2020)
%doi:10.1103/PhysRevLett.125.181603
[arXiv:2005.13897 [hep-th]].
%11 citations counted in INSPIRE as of 15 Jan 2021

\bibitem{PMTidal}
D.~Bini, T.~Damour and A.~Geralico,
%``Scattering of tidally interacting bodies in post-Minkowskian gravity,''
Phys. Rev. D \textbf{101}, no.4, 044039 (2020)
%doi:10.1103/PhysRevD.101.044039
[arXiv:2001.00352 [gr-qc]];
%
K.~Haddad and A.~Helset,
%``Gravitational tidal effects in quantum field theory,''
[arXiv:2008.04920 [hep-th]];
%
R.~Aoude, K.~Haddad and A.~Helset,
%``Tidal effects for spinning particles,''
[arXiv:2012.05256 [hep-th]].
%1 citations counted in INSPIRE as of 02 Jan 2021

\bibitem{PMRadiation}
A.~Luna, R.~Monteiro, I.~Nicholson, D.~O'Connell and C.~D.~White,
%``The double copy: Bremsstrahlung and accelerating black holes,''
JHEP \textbf{06}, 023 (2016)
%doi:10.1007/JHEP06(2016)023
[arXiv:1603.05737 [hep-th]];
%115 citations counted in INSPIRE as of 15 Jan 2021
%
W.~D.~Goldberger and A.~K.~Ridgway,
%``Radiation and the classical double copy for color charges,''
Phys.\ Rev.\ D {\bf 95}, no. 12, 125010 (2017)
%doi:10.1103/PhysRevD.95.125010
[arXiv:1611.03493 [hep-th]];
%%CITATION = doi:10.1103/PhysRevD.95.125010;%%
%75 citations counted in INSPIRE as of 25 Oct 2019
%
W.~D.~Goldberger, S.~G.~Prabhu and J.~O.~Thompson,
%``Classical gluon and graviton radiation from the bi-adjoint scalar double copy,''
Phys.\ Rev.\ D {\bf 96}, no. 6, 065009 (2017)
%doi:10.1103/PhysRevD.96.065009
[arXiv:1705.09263 [hep-th]];
%%CITATION = doi:10.1103/PhysRevD.96.065009;%%
%51 citations counted in INSPIRE as of 25 Oct 2019
%
W.~D.~Goldberger and A.~K.~Ridgway,
%``Bound states and the classical double copy,''
Phys.\ Rev.\ D {\bf 97}, no. 8, 085019 (2018)
%doi:10.1103/PhysRevD.97.085019
[arXiv:1711.09493 [hep-th]];
%%CITATION = doi:10.1103/PhysRevD.97.085019;%%
%34 citations counted in INSPIRE as of 25 Oct 2019
%
D.~Chester,
%``Radiative double copy for Einstein-Yang-Mills theory,''
Phys.\ Rev.\ D {\bf 97}, no. 8, 084025 (2018)
%doi:10.1103/PhysRevD.97.084025
[arXiv:1712.08684 [hep-th]];
%%CITATION = doi:10.1103/PhysRevD.97.084025;%%
%20 citations counted in INSPIRE as of 25 Oct 2019
%
W.~D.~Goldberger, J.~Li and S.~G.~Prabhu,
%``Spinning particles, axion radiation, and the classical double copy,''
Phys.\ Rev.\ D {\bf 97}, no. 10, 105018 (2018)
%doi:10.1103/PhysRevD.97.105018
[arXiv:1712.09250 [hep-th]];
%%CITATION = doi:10.1103/PhysRevD.97.105018;%%
%37 citations counted in INSPIRE as of 25 Oct 2019
%
C.~H.~Shen,
%``Gravitational radiation from color-kinematics duality,''
JHEP {\bf 1811}, 162 (2018)
%doi:10.1007/JHEP11(2018)162
[arXiv:1806.07388 [hep-th]];
%%CITATION = doi:10.1007/JHEP11(2018)162;%%
%28 citations counted in INSPIRE as of 25 Oct 2019
%
Y.~F.~Bautista and A.~Guevara,
%``On the double copy for spinning matter,''
[arXiv:1908.11349 [hep-th]];
%%CITATION = ARXIV:1908.11349;%%
%2 citations counted in INSPIRE as of 25 Oct 2019
%
W.~D.~Goldberger and J.~Li,
%``Strings, extended objects, and the classical double copy,''
JHEP \textbf{02}, 092 (2020)
%doi:10.1007/JHEP02(2020)092
[arXiv:1912.01650 [hep-th]];
%21 citations counted in INSPIRE as of 15 Jan 2021
%
G.~L.~Almeida, S.~Foffa and R.~Sturani,
%``Classical Gravitational Self-Energy from Double Copy,''
JHEP \textbf{11}, 165 (2020)
%doi:10.1007/JHEP11(2020)165
[arXiv:2008.06195 [gr-qc]].
%6 citations counted in INSPIRE as of 15 Jan 2021

\bibitem{SoftRadiation}
A.~Laddha and A.~Sen,
%``Gravity waves from soft theorem in general dimensions,''
JHEP {\bf 1809}, 105 (2018)
%doi:10.1007/JHEP09(2018)105
[arXiv:1801.07719 [hep-th]];
%%CITATION = doi:10.1007/JHEP09(2018)105;%%
%
A.~Laddha and A.~Sen,
%``Logarithmic terms in the soft expansion in four dimensions,''
JHEP {\bf 1810}, 056 (2018)
%doi:10.1007/JHEP10(2018)056
[arXiv:1804.09193 [hep-th]];
%%CITATION = doi:10.1007/JHEP10(2018)056;%%
%
A.~Laddha and A.~Sen,
%``Observational Signature of the Logarithmic Terms in the Soft Graviton Theorem,''
Phys. Rev. D \textbf{100}, no.2, 024009 (2019)
%doi:10.1103/PhysRevD.100.024009
[arXiv:1806.01872 [hep-th]];
%29 citations counted in INSPIRE as of 15 Jan 2021
%
B.~Sahoo and A.~Sen,
%``Classical and Quantum Results on Logarithmic Terms in the Soft Theorem in Four Dimensions,''
JHEP \textbf{02}, 086 (2019)
%doi:10.1007/JHEP02(2019)086
[arXiv:1808.03288 [hep-th]];
%46 citations counted in INSPIRE as of 15 Jan 2021
%
M.~Ciafaloni, D.~Colferai and G.~Veneziano,
%``Infrared features of gravitational scattering and radiation in the eikonal approach,''
Phys. Rev. D \textbf{99}, no.6, 066008 (2019)
%doi:10.1103/PhysRevD.99.066008
[arXiv:1812.08137 [hep-th]];
%29 citations counted in INSPIRE as of 15 Jan 2021
%
A.~Laddha and A.~Sen,
%``Classical proof of the classical soft graviton theorem in D\ensuremath{>}4,''
Phys. Rev. D \textbf{101}, no.8, 084011 (2020)
%doi:10.1103/PhysRevD.101.084011
[arXiv:1906.08288 [gr-qc]];
%16 citations counted in INSPIRE as of 15 Jan 2021
%
A.~P.V. and A.~Manu,
%``Classical double copy from Color Kinematics duality: A proof in the soft limit,''
Phys. Rev. D \textbf{101}, no.4, 046014 (2020)
%doi:10.1103/PhysRevD.101.046014
[arXiv:1907.10021 [hep-th]];
%18 citations counted in INSPIRE as of 15 Jan 2021
%
A.~P.~Saha, B.~Sahoo and A.~Sen,
%``Proof of the classical soft graviton theorem in $D$ = 4,''
JHEP \textbf{06}, 153 (2020)
%doi:10.1007/JHEP06(2020)153
[arXiv:1912.06413 [hep-th]];
%16 citations counted in INSPIRE as of 15 Jan 2021
%
M.~A, D.~Ghosh, A.~Laddha and P.~V.~Athira,
%``Soft Radiation from Scattering Amplitudes Revisited,''
[arXiv:2007.02077 [hep-th]];
%9 citations counted in INSPIRE as of 15 Jan 2021
%
B.~Sahoo,
%``Classical Sub-subleading Soft Photon and Soft Graviton Theorems in Four Spacetime Dimensions,''
JHEP \textbf{12}, 070 (2020)
%doi:10.1007/JHEP12(2020)070
[arXiv:2008.04376 [hep-th]].
%9 citations counted in INSPIRE as of 15 Jan 2021

\end{thebibliography}
\end{document}